\documentclass[amsmath,amssymb,aps]{revtex4-2}

\usepackage{graphicx}
\usepackage{dcolumn}
\usepackage{bm}

\usepackage{xcolor}

\begin{document}

\preprint{APS/123-QED}

\title{Unsupervised Learning of Quantum Phase Transitions for Bose-Hubbard lattice systems}

\author{Bihui Zhu}

\affiliation{%
 Homer L. Dodge Department of Physics and Astronomy,
The University of Oklahoma, Norman, Oklahoma 73019, USA\\
 Center for Quantum Research and Technology, The University of Oklahoma, Norman, Oklahoma 73019, USA
}%
 \email{bihui.zhu@ou.edu}


\begin{abstract}
{\bf Abstract}: 
Characterizing quantum many-body phase structure is a major goal for quantum simulation. Here, we employ an unsupervised learning approach based on diffusion maps to learn phase transitions in bosonic lattice systems described by Bose–Hubbard type models, which can be realized in ultracold atoms and related quantum simulation platforms. We demonstrate that this approach identifies phase structure across distinct settings without prior knowledge of order parameters or handcrafted observables, including ground-state transitions involving symmetry-protected topological phases and nonequilibrium regimes distinguishing ergodic and many-body localized behavior. Our results indicate that the approach has the potential for direct application to experimentally accessible measurement data for learning quantum phases in current quantum simulators.

\end{abstract}
\maketitle


\section{\label{sec:level1}Introduction}
Understanding quantum many-body phase transitions is a central objective of condensed matter physics and a primary application of programmable quantum simulators. Toward this goal, recent advances in ultracold atoms, superconducting circuits, and related platforms now enable the preparation and measurement of strongly correlated lattice systems with unprecedented control, opening new opportunities to probe symmetry breaking, topology, and ergodicity in interacting quantum matter \cite{Gross2017Science,Vuletic2021PRXQuantum,Bernien2017Nature51Atom,Bluvstein2021ScienceDrivenRydberg,Ebadi2021Nature256Atom, Semeghini2021ScienceSpinLiquid, Manovitz2025NatureCoarsening, Martinez2016NatureSchwinger, Smith2016NatPhysMBL,Zhang2017NatureDynamicalPT, Satzinger2021ScienceToricCode, Andersen2025NatureAnalogueDigital, Mi2022NatureTimeCrystallineEigenstate,lebrat2024observation,leonard2023realization,xu2023frustration}.

A paradigmatic setting for exploring such many-body phenomena is provided by the Bose–Hubbard model (BHM), which describes interacting bosons hopping on a lattice and can be realized with high tunability in ultracold atomic gases in optical lattices \cite{Fisher1989PRB,Jaksch1998PRL,greiner2002quantum,Bloch2008RMP,Gross2017Science}. 
Modern quantum-gas microscope experiments provide single-site-resolved measurement snapshots and enable direct access to microscopic configurations in bosonic lattice systems, including recent studies of dynamical  regimes and topological states beyond the textbook superfluid--Mott setting \cite{bakr2009quantum,sherson2010single,Gross2021ARCM, Wei2022Science,QGMBoseGlass2025arXiv, suTopologicalPhaseTransitions2025}.
Closely related Hamiltonians can also be  implemented in superconducting circuit platforms, where photonic excitations in coupled resonators with Josephson nonlinearities naturally realize Bose–Hubbard–type dynamics \cite{Ma2019Nature,Fitzpatrick2017PRX,Zhang2023Science,Du2024PRL,Du2025PRR}.  The BHM and its extensions host a range of interaction-driven phases in equilibrium and dynamical regimes whose characterization remains nontrivial. In addition, in comparison to qubit systems with fixed two-level local structure, bosonic lattice sites possess multi-level occupation degrees of freedom, leading to a more rapidly growing state space. The exploration of Bose Hubbard physics has been of significant experimental interest across platforms \cite{Gross2017Science,kim2025multi,Wei2023PRXBraneParity,blochPrethermalPRX2020,Ma2019Nature,Zhang2023Science}.

At the same time, characterizing quantum phases from experimental output requires suitable diagnostic approaches. 
In traditional analyses, phase classification typically proceeds by constructing appropriate phase-indicator observables from the measurement data. For example, for a simple symmetry-breaking phase whose order parameter is a known local observable, evaluating this quantity while sweeping across system parameters provides a standard approach to demonstrating a phase transition in quantum simulation. However, for more complex quantum phases, the relevant diagnostics may involve quantities that are experimentally challenging or computationally costly to assemble, such as nonlocal string order parameters or entanglement-based measures \cite{Hasan2010RMP,Qi2011RMP,Lukin2019Science,kaufman2016quantum,abaninColloquiumManybodyLocalization2019,chernnumberHafeziPRL2021,elben2020many}. In addition, when exploring new models or studying new parameter regimes, there may be insufficient prior insight to guide the choice of the proper diagnostic observable or feature to extract from the measurements. These  motivate the development of data-driven approaches capable of inferring phase structure efficiently from accessible measurement output.

Machine learning methods have become widely integrated into the study of quantum many-body systems. In the context of phase identification, a significant body of work built on supervised approaches has been developed, such as neural-network classifiers \cite{Carrasquilla2017NatPhys,vanNieuwenburg2017NatPhys, phaseNNexpNatPhys2019}. In such supervised approaches, the ML model is trained with labeled data and learns to identify phases with new parameters. To discover new phases and reveal unknown phase transitions, unsupervised learning approaches are very helpful as they do not require prior knowledge of the phases or preassigned labels. These encompass 
principal component analysis (PCA), autoencoder-based representations, and nonlinear manifold-learning approaches \cite{Wang2016PRB,Hu2017PRE,classicalAE2017,AEexp2021,chngUnsupervisedMachineLearning2018a,rodriguez-nievaIdentifyingTopologicalOrder2019,lidiakUnsupervisedMachineLearning2020}.  
Among these, diffusion-map--based manifold learning has been shown to be effective for extracting phase structure; nevertheless, many existing implementations rely on engineered features derived from prior physical insight, or require carefully tailored preprocessing pipelines, or have primarily focused on spin or fermionic lattice systems \cite{cheTopologicalQuantumPhase2020a,yuUnsupervisedLearningInteracting2024,zivUnsupervisedMachineLearning2025, rodriguez-nievaIdentifyingTopologicalOrder2019,lidiakUnsupervisedMachineLearning2020,kuoUnsupervisedLearningInteracting2022}


In this work, we provide an unsupervised learning framework based on diffusion maps (DM) tailored to Bose-Hubbard type models relevant to recent quantum simulator experiments, operating directly on measurement snapshots. 
We show that our approach can capture a variety of quantum many-body phases hosted in these systems both in equilibrium and nonequilibrium settings, without predefined order parameters or engineered preprocessing of the experimental output. The DM algorithm employed here is computationally efficient for systematically exploring the phase diagram as it does not require joint processing of datasets across all model parameters. Our results establish this approach as a practical method for characterizing phase structure in bosonic lattice quantum simulators.

\section{\label{sec:model}Model and method}
\subsection{Bosonic particles in a lattice}\label{sec:BHM}
The Bose Hubbard model realized in many quantum simulation experiments  is  described by 
\begin{eqnarray}
\hat H=-J\sum_i{ (\hat a_i^\dagger\hat a_{i+1}+h.c.})+\frac{U}{2}\sum_i\hat n_i(\hat n_i-1)+\sum_{i,r}V_r\hat n_i\hat n_{i+r}+\sum_i\mu_i\hat n_i, \label{eq:BHM}
\end{eqnarray}
where $J$ denotes the tunneling rate between neighboring site, $U$ is the on-site interaction strength, $\mu_i$ is the local chemical potential, and $V_r$ accounts for interactions between atoms occupying different sites, which naturally arise in bosonic quantum simulators with dipolar particles, such as ultracold Er and Cr atoms \cite{ebhmFrancescaSci2016,ferstererDynamicsItinerantSpin32019,suDipolarQuantumSolids2023}. Across this work, we use $J$ as the energy unit and set $J=1$. 
The BHM provides a prototypical description of interacting bosons on a lattice, capturing the competition between the  tunneling, which delocalize partiles across the lattice, and the on-site interactions, which disfavors having more than one particle per site.

\begin{figure*}
\centering
\includegraphics[width=0.6\textwidth]{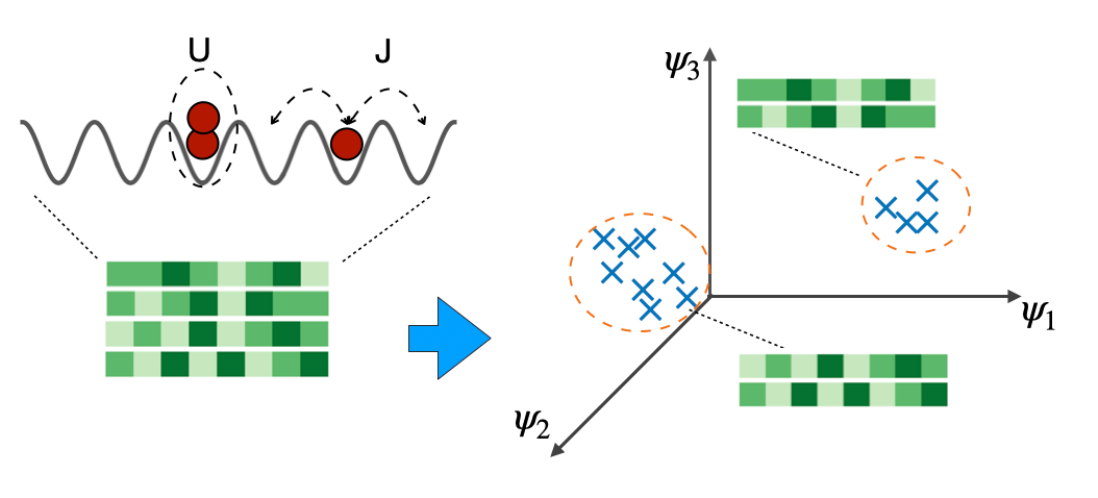}
\caption{Schematic illustration of the application of diffusion map approach for learning quantum phase transitions. Measurement of onsite boson occupations is read out from quantum simulator experiment or numerical simulation. The  measurement snapshots are collected for each set of model parameters $U$, $J$, etc. and the diffusion map identifies similar configurations in the measurement snapshots through a diffusion process, which reveals the underlying structure in different phases and enables unsupervised phase classification.}\label{fig1}
\end{figure*}
\subsection{Diffusion maps}\label{sec:DM}
Diffusion maps is a kernel-based manifold learning method that constructs a diffusion process on the dataset. By specifying transition probabilities from one configuration to another based on a similarity kernel, the method induces a random walk in the sample space. Repeated random walks across the sample space allow exploring the global structure of the dataset that is not apparent from direct pairwise distances alone. Compared to conventional linear dimensional reduction methods such as PCA, diffusion maps can reveal nonlinear structures that are important for identifying complex phases.   

For a dataset with $M$ samples, 
\[
X = \{\mathbf{x}_i\}_{i=1}^{M}, 
\qquad 
\mathbf{x}_i = \left(x_i^{(1)}, x_i^{(2)}, \ldots, x_i^{(L)}\right),
\]
where each $\mathbf{x}_i \in \mathbb{R}^L$ is an $L$-dimensional vector, we construct a Gaussian kernel 
\begin{equation}
    K_{ij}=\mathrm{exp}\left( -\frac{d_{ij}^2}{2\epsilon}\right).
\end{equation}
Here, $d_{ij}$ is the distance between samples and in this work we use the Euclidean distance metric $d_{ij}^2=\sum_{k}(x_i^{(k)}-x_j^{(k)})^2/\mathcal{N}$, normalized such that $d_{ij}\in[0,1]$. The hyperparameter $\epsilon$ sets how rapidly the kernel weights decay with distance and determines the local connectivity among samples, and thus controls how strongly the diffusion process mixes configurations in one step. From this, the row-normalized matrix $P$ is constructed  with  
\begin{equation}
P_{ij}=\frac{K_{ij}}{\sum_j K_{ij}}.
\end{equation}
The transition matrix $P$ defines the diffusion on the sample space, with its element $P_{ij}$ denoting the one-step transition probability from sample $i$ to $j$. The eigenvalues and eigenvectors of the transition matrix, denoted by $\lambda_k$ and $\Psi_k$, determine the diffusion distance between samples at diffusion time $t$, with each eigenmode contributing with weight $\lambda_k^t$. Eigenmodes with eigenvalues close to 1 are long-lived and encode the global structure of the dataset under diffusion. The number of eigenvalues larger than $1-\delta$ then provides an estimate of the number of dominant long-lived components (clusters) that persist beyond a timescale of order $\sim 1/\delta$. By fixing the threshold $\delta$ and choosing the scale $\epsilon$ appropriately, the resulting cluster count can reveal changes in the underlying phase structure as the Hamiltonian parameters are varied. The hyperparameter $\epsilon$ is associated with the scale at which differences between configurations of distinct clusters are resolved, which depends on the actual distance distribution of the dataset. In practice, $\epsilon$ is scanned over a range to obtain the value where the cluster count shows a conspicuous change across datasets \cite{lidiakUnsupervisedMachineLearning2020,kuoUnsupervisedLearningInteracting2022}.

In this work, we apply the above DM approach to study the phases in three variants of the BHM, motivated by their close relevance to cold-atom quantum simulator experiments. In each case, we generate synthetic data from numerical simulations of the corresponding model using 
standard numerical methods, such as exact diagonalization or density matrix renormalization group (DMRG). More concretely, for each set of Hamiltonian parameters $\boldsymbol{\theta}$, we obtain the quantum state wavefunction $\psi({\boldsymbol \theta})$ from the solving the ground state or dynamical evolution under the Hamiltonian. From the wavefunction, measurement snapshots are generated by sampling in the onsite occupation number basis. That is, for each shot, we perform a projective measurement of the local number operator $\hat n_i$ on all $L$ lattice sites, drawing configurations ${\bf n}=(n_1,n_2,...,n_L)$ according to the Born probabilities 
\begin{eqnarray}
    P({\bf n})=|\langle {\bf n}|\psi ({\boldsymbol \theta})\rangle|^2.
\end{eqnarray}
For a maximum number of $n_{\rm max}$ bosons per site, each $n_l$ can take integer values in $[0,n_{\rm nmax}]$, and each sampled configurations automatically satisfies $\sum_l n_l=N$, with the total boson number $N=\overline{n}L$ fixed at a given average filling fraction $\overline{n}$. The configuration ${\bf n}$ is used to represent each sample ${\bf x}$, and for each $\boldsymbol{\theta}$, $M$  snapshots are generated independently. We use $M=500$ and fixed $\delta=10^{-2.5}$ across all the applications presented.

In these studies, we aim to demonstrate the capacity of the DM approach and the calculations are carried out for one-dimensional geometry and modest system sizes, for which the snapshot datasets can be generated efficiently using numerical simulations. Meanwhile, as described above, the DM approach operates directly on measurement samples and is thus expected to remain applicable to systems in higher spatial dimensions and of larger sizes when the corresponding measurement data are available. We note that the DM approach studied here aims to learn phase structure from measurement data without requiring prior knowledge of phase labels or phase-specific diagnostic quantities, 
rather than to provide a high-precision determination of critical points.

\section{\label{sec:results} Results}
\subsection{Phases in an extended Bose Hubbard model}\label{sec:EBHM}

We first study the phase diagram of the extended BHM with nearest neighbor interactions, setting $V_1=V$ with $V_{r>1}=0$ and $\mu_i=0$
in Eq. (\ref{eq:BHM}). Namely, 
\begin{eqnarray}
\hat H=-J\sum_i{ (\hat  a_i^\dagger\hat a_{i+1}+h.c.})+\frac{U}{2}\sum_i\hat n_i(\hat n_i-1)+V\sum_{i}\hat n_i\hat n_{i+1}, \label{eq:model1}
\end{eqnarray}
The presence of inter-site interactions $V$ can enrich the phases beyond the BHM with only on-site interactions, giving rise to a topologically nontrivial phase with hidden order in the fluctuations, which carries analogies with the Haldane phase in integer spin chains \cite{dallatorreHiddenOrder1D2006, ejimaSpectralEntanglementProperties2014}. This Haldane insulator (HI) phase is characterized by the nonlocal string correlation $O_s=\mathrm{\lim}_{|i-j|\rightarrow \infty}\langle \delta \hat n_i \mathrm{ exp}(i\pi\sum_{i\le k<j}\delta \hat n_k)\delta \hat n_j\rangle\rightarrow\mathrm{const}\neq 0$, and is distinguished from the Mott insulator (MI) phase and the density-wave (DW) phase \cite{bergRiseFallHidden2008}. Measuring such long string correlators is usually challenging for experiments, and therefore alternative experimentally feasible probes are useful.  

We focus on the insulating phase regimes where interaction strengths are not small, and vary the interaction strengths $U$ and $V$ while keeping the tunneling rate $J=1$ fixed and the average filling $\overline{n}=1$, as in early studies on such an extended Hubbard model \cite{dallatorreHiddenOrder1D2006, rossiniPhaseDiagramExtended2012, ejimaSpectralEntanglementProperties2014}. The emergence of the HI phase can be understood via the interplay of the delocalization from tunneling $J$ and the density ordering from the interaction $V$. As such, the HI phase is realized at intermediate $V/J$, with MI phase realized when the on-site interactions $U$ become dominant and the DW phase when $V$ dominates \cite{dallatorreHiddenOrder1D2006, rossiniPhaseDiagramExtended2012}.

\begin{figure}[h]
\centering
\includegraphics[width=0.45\textwidth]{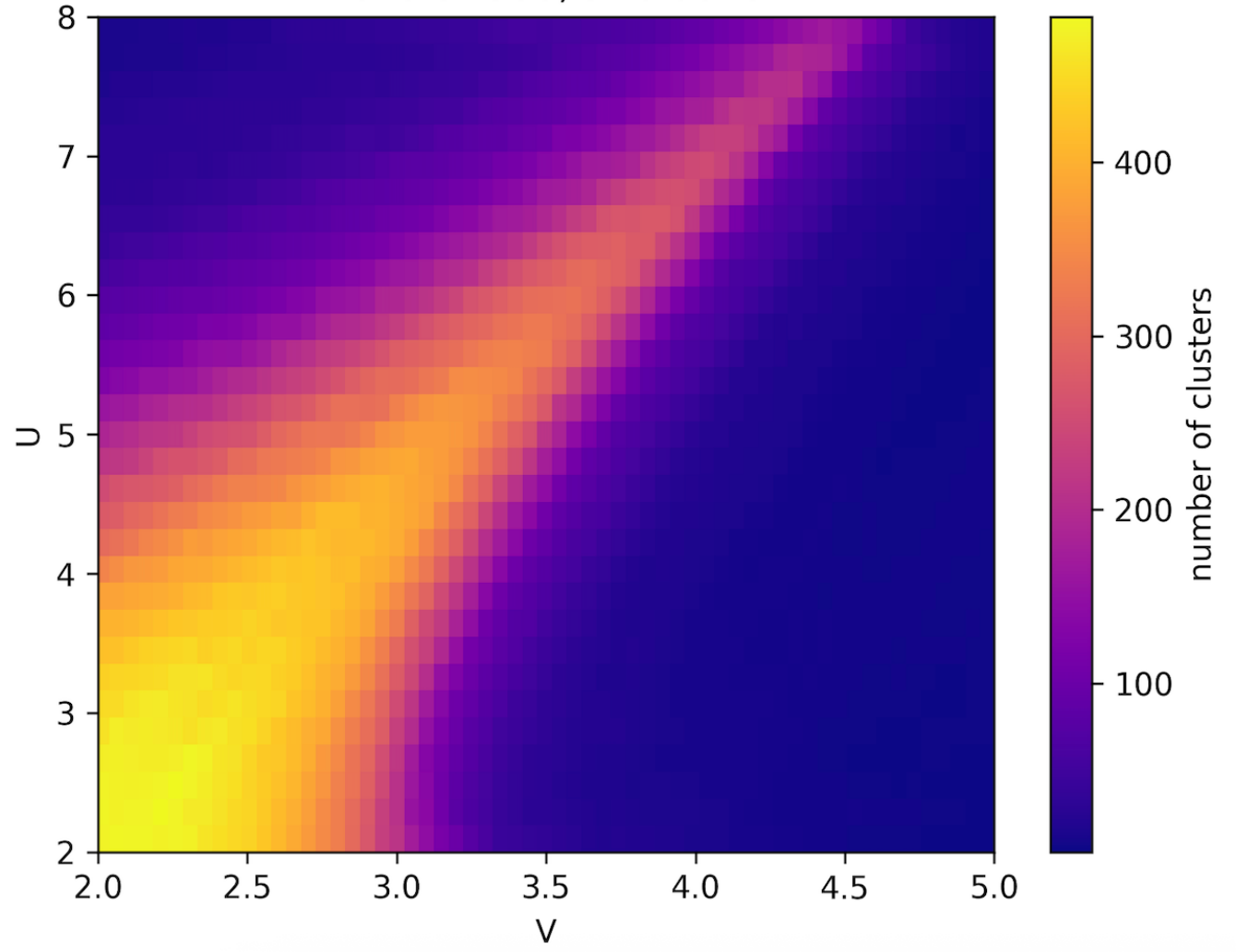}
\caption{Ground state phase diagram for the 1D extended Bose Hubbard model at unit filling. The color bar represents the number of clusters obtained from the DM approach.}\label{fig:EBHM}
\end{figure}
To apply the DM approach, we use DMRG to obtain the ground state of the extended Hubbard model Eq.~(\ref{eq:model1}) in 1D with a maximum of 3 bosons per site over a parameter grid $(U,V)$. From the ground state wavefunction, we then sample the measurement shots of the on-site population in the lattice. The DM is implemented on the measurement shots captured for each set of parameter $(U,V)$. The numbers of clusters obtained from the DM across the parameter grid are plotted in Fig. \ref{fig:EBHM}, where $L=16$ sites and 500 samples of shots are used for each parameter point with fixed $\epsilon=0.0018$. The DM result clearly separates the parameter space into three phase regimes, leading to a phase diagram akin to  the results in Ref. \cite{rossiniPhaseDiagramExtended2012}. For Hamiltonian parameters with comparable $U$ and $V$, there is a conspicuous indication of a distinct phase, corresponding to the HI phase in the extended BHM.

\subsection{Learning the phase transition between crystalline SPT phases}\label{sec:greiner}
As a second application, we study the case with a large on-site interaction $U$ and an additional staggered chemical potential, motivated by its direct relevance to recent quantum gas microscope experiments using ultracold Er atoms \cite{suTopologicalPhaseTransitions2025}. Concretely, the system is described by 
\begin{eqnarray}
\hat H=-J\sum_i{(\hat  a_i^\dagger\hat a_{i+1}+h.c.})+\frac{U}{2}\sum_i\hat n_i(\hat n_i-1)+\sum_{i,r}V_{\rm dip}(r)\hat n_i\hat n_{i+r}+\mu_s\sum_i(-1)^{i}\hat n_i, \label{eq:greiner}
\end{eqnarray}
where the inter-site interaction $V_{\rm dip}(r)$ arises from the dipolar interactions between magnetic Er atoms. This model hosts  a quantum phase transition between two crystalline symmetry protected topological phases (CSPTs) in 1D at unit filling $\overline{n}=1$, protected by the  symmetries from a global parity and a site-centered inversion symmetry \cite{suTopologicalPhaseTransitions2025,fuTopologicalCrystallineInsulators2011,hughesInversionsymmetricTopologicalInsulators2011,turnerEntanglementInversionSymmetry2010}. 
The SPT phases, CSPT-1 and CSPT-2, are realized at weak and strong staggered chemical potentials, respectively, and cannot be adiabatically connected while preserving the protecting symmetries \cite{suTopologicalPhaseTransitions2025}. 
In particular, they cannot be distinguished by local order parameters but a nonlocal string operator needs to measured to identify these phases, which has the form of $\hat{P}_{j}(r)=\prod_{j\le k<j+r} \mathrm{exp}({i\pi\hat n_k})$. The corresponding string order parameters $p_1$ and $p_2$ for identifying the two phases are obtained from averaging $(-1)^rP_j(r)$ and $P_j(r)$ over long strings, which are finite in one phase and vanish in the other phase.

\begin{figure}[h]
\centering
\includegraphics[width=0.45\textwidth]{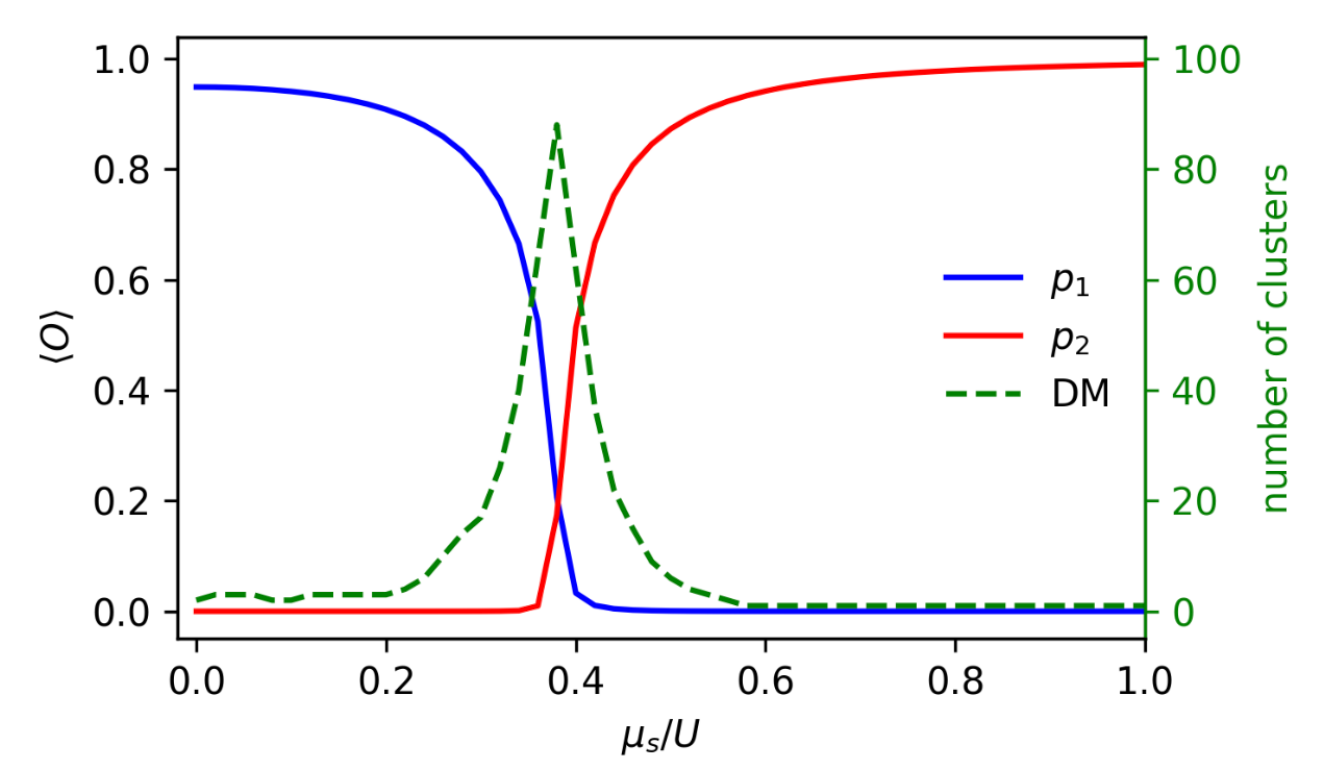}
\caption{Quantum phase transition between two crystalline SPTs in model Eq.~(\ref{eq:greiner}). The blue (red) line is the string order parameter $p_1$ ($p_2$) obtained directly from DMRG calculations, which indicates the system changes between two topologically distinct phases. The green dashed line shows the result from the DM approach.} \label{fig:greiner}
\end{figure}
We use DMRG to find the ground state of this system at unit filling, for a chain with $L=16$ sites, held at $U/J=20$ and $V_{\rm nn}/J=2.5$, as relevant to experimental settings in Ref. \cite{suTopologicalPhaseTransitions2025}. In our calculations, the maximum occupation per site used is 3, and for simplicity,  we have truncated the dipolar interactions to nearest neighbors with strength $V_{\rm nn}$. With the measurement shots obtained from sampling the DMRG ground state, we apply the DM approach for a range of values or the staggered chemical potential magnitude $\mu_s$, with $500$ samples for each parameter point and $\epsilon=0.001$ fixed. The results as shown in Fig. \ref{fig:greiner} demonstrate a pronounced change around intermediate $\mu_s$, indicating a transition between two separated phases.

As a comparison, we also compute the string order parameters $p_1$ and $p_2$ from the DMRG calculations, which are plotted together in Fig.  \ref{fig:greiner}. The string order parameter $p_1$ is significant for small $\mu_s$, where the system is in the CSPT-1 phase, and it vanishes above a threshold value of $\mu_s$. Meanwhile, $p_2$ vanishes in CSPT-1 phase, and becomes significant as the magnitude of the staggered chemical potential $\mu_s$ is varied cross the quantum critical point and the system enters the CSPT-2 phase. The DM result shows a remarkably good agreement with the transition indicated by $p_1$ and $p_2$, demonstrating the DM's capability to capture the quantum phase transition between two SPTs in this model. Moreover, while the sharp peak in the DM result can provide an indication of the phase transition, it is worth noting that a visible asymmetry is observed between the small $\mu_s$ regime and the large $\mu_s$ regime. In particular, for large $\mu_s$, the DM result remains smaller than that for small $\mu_s$. Such an asymmetry suggests the DM identifies different structures in the underlying CSPT-1 and CSPT-2 phases, which are topologically distinct, aside from responding to the critical point in this system.

\subsection{Application for many-body localization}
The DM approach can be applied beyond classifying ground state phases. Here, we use DM to study the out-of-equilibrium phenomena in a disordered lattice boson system, where the system dynamics may change from thermal to a Many-body localization (MBL) phase in the presence of strong disorder \cite{rispoliQuantumCriticalBehaviour2019}. The MBL phase transition exhibits properties fundamentally different from the ground state phase transitions and has been the subject of significant research interest both experimentally and theoretically \cite{baskoMetalInsulatorTransition2006, abaninRecentProgressManybody2017, nandkishoreManyBodyLocalizationThermalization2015, husePhenomenologyFullyManybodylocalized2014, wahlSignaturesManybodyLocalized2019, aletManybodyLocalizationIntroduction2018, sierantManybodyLocalizationBosons2018, yaoManybodyLocalizationBoseHubbard2020,luschenObservationSlowDynamics2017,schreiberObservationManybodyLocalization2015,choiExploringManybodyLocalization2016,kondovDisorderInducedLocalizationStrongly2015, rispoliQuantumCriticalBehaviour2019}. In particular, such a phase transition is not identified via simple order parameters but manifests through quantities such as level statistics or entanglement features, which are difficult to measure experimentally. Meanwhile, fully resolving the level statistics and entanglement properties is challenging for numerical calculations to extend to large system sizes, and many studies have focused on spin-1/2 or fermionic systems where the local Hilbert space size is smaller \cite{abaninRecentProgressManybody2017, nandkishoreManyBodyLocalizationThermalization2015, husePhenomenologyFullyManybodylocalized2014, wahlSignaturesManybodyLocalized2019,palManybodyLocalizationPhase2010,khemaniCriticalPropertiesManyBody2017}.  

\begin{figure}[h]
\centering
\includegraphics[width=0.8\textwidth]{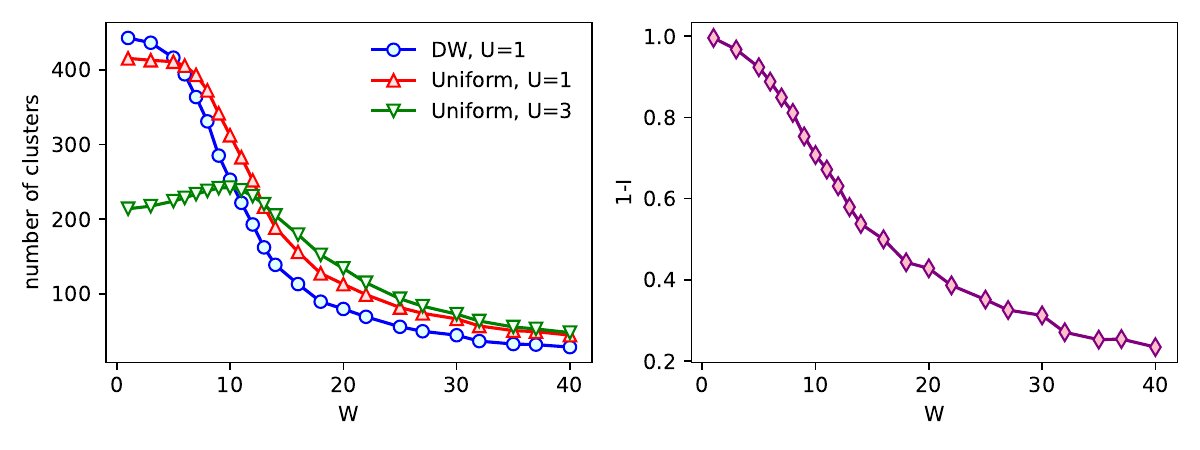}
\caption{Dynamical phases in a disordered Bose Hubbard chain. Left: DM results for quench dynamics from the density-wave initial state (blue), and uniform initial states at different on-site interaction strengths $U=1$ (red) and $U=3$ (green). Right: The imbalance $I$ is measured at the end of the dynamics, and $1-I$ is plotted. For these results, $L=8$ with $\overline{n}=1$ and a maximum number of 4 bosons per site. The system is evolved until the time $t=200/J$. }\label{fig:mbl}
\end{figure}
Here, we probe the MBL by using measurement shots obtained from the quench dynamics, a procedure relevant to many cold atom experiments \cite{schreiberObservationManybodyLocalization2015,luschenObservationSlowDynamics2017,rispoliQuantumCriticalBehaviour2019, zivUnsupervisedMachineLearning2025}. In such experiments, the system is prepared in some simple, typically structured initial state, and undergoes time evolution governed by the disordered Hamiltonian for a long time, and then the atom numbers in the final state are read out. 
Concretely, we consider the model Hamiltonian

\begin{eqnarray}
\hat H=-J\sum_i{ (\hat a_i^\dagger\hat a_{i+1}+h.c.})+\frac{U}{2}\sum_i\hat n_i(\hat n_i-1)+\sum_i\mu_i\hat n_i, \label{eq:MBL}
\end{eqnarray}
where $\mu_i$ is the random on-site potential, uniformly distributed within $[-W/2, W/2]$ and $W$ denotes the disorder strength.

To obtain the quantum state evolved under Eq.~(\ref{eq:MBL}) after a long time, we use exact diagonalization for a 1D chain at unit filling and a maximum of 4 bosons per site. We first consider the dynamics starting from an initial state with the DW pattern $|2020...20\rangle$ with on-site interaction $U=1$ and $L=8$ sites, a setting adopted in the exact diagonalization studies in Ref. \cite{yaoManybodyLocalizationBoseHubbard2020}. For each realization of randomly sampled chemical potential $\mu_i$, we evolve the system until $Jt=200$--a typical time accessible in cold-atom experiments \cite{rispoliQuantumCriticalBehaviour2019}.
From the final quantum state obtained, we sample 500 measurement shots of the on-site atom population and apply the DM approach at fixed $\epsilon=0.001$ (more details provided in  Appendix \ref{sec:appendix}). The DM results calculated across a range of disorder magnitude $W$ are shown in Fig. \ref{fig:mbl} (blue), where for each $W$ the result is averaged over 500 disorder realizations. The DM results demonstrate a clear difference between the weak-disorder and strong-disorder regimes, suggesting the presence of a thermal phase and an MBL phase. In the latter, the sample configurations remain similar to the initial state due to localization, resulting in a small cluster count.

For this initial DW state, the imbalance 
\begin{eqnarray}
    I=\frac{N_e-N_o}{N_e+N_o}
\end{eqnarray}
if often used as an indicator for the relaxation dynamics \cite{schreiberObservationManybodyLocalization2015,luschenObservationSlowDynamics2017}. In the thermal phase, the imbalance vanishes in the long time, and a finite stationary value of the imbalance indicates localized dynamics.  To further confirm the onset of MBL, we also compute the imbalance from the same exact diagonalization data, and averaged the result over 500 disorder realizations for each $W$. In the right panel of Fig. \ref{fig:mbl}, we plot $1-I$ obtained at the end of the long time dynamics. Due to the finite system size and evolution time, the transition in the imbalance is not sharp, while a clear variation in the magnitude of the final imbalance is evident, indicating thermalizing dynamics for small $W$ and localizing behavior for large $W$. A comparison of the two panels shows that the trend observed in the DM results is similar to that revealed by the imbalance, supporting the DM's ability to probe MBL phase in this system.

As MBL is an eigenstate property, it can occur for different initial states. Prior studies have shown the existence of a many-body mobility edge, such that the critical disorder strength for MBL can depend on the initial state's energy density \cite{luitzManybodyLocalizationEdge2015,sierantManybodyLocalizationBosons2018,guoObservationEnergyresolvedManybody2021}. This dependence has been reported to be more pronounced at stronger interactions, while for weaker interactions the dependence is weaker in the Bose Hubbard system \cite{yaoManybodyLocalizationBoseHubbard2020,sierantManybodyLocalizationBosons2018}. Accordingly, we also study dynamics from the initial state with uniform density  $|111...1\rangle$, which can be easily prepared in many cold-atom experiments. From the level statistics analysis, the  behavior for such uniform initial state has been found to be similar to the DW initial state at $U=1$  \cite{yaoManybodyLocalizationBoseHubbard2020}. Meanwhile, the imbalance is identically zero for the uniform initial state and cannot serve as a diagnostic as in the DW initial state case, making the detection of MBL more challenging experimentally. Nevertheless,  the measurement shots obtained after time-evolution can be used to run the DM method with the same procedure as for the DW initial state case (with $\epsilon=0.001$ fixed). The DM result is shown in Fig. \ref{fig:mbl} (red), which indicates a change in behavior between the weak and disorder parameter regimes similar to the DW initial state. 

As another test for the method, we also apply DM to the dynamics  evolved from the  uniform initial state at a stronger interaction strength $U=3$ and the result is plotted together in Fig. \ref{fig:mbl} (green). In this case, there is a visible difference from the results obtained at $U=1$, with the indicated transition occurring at a higher critical disorder strength $W$, consistent with previous theoretical predictions \cite{yaoManybodyLocalizationBoseHubbard2020,orellProbingManybodyLocalization2019}. In the Appendix, we also compare the DM results  between different initial states for the strong interaction case, where clearly distinguishable behavior is observed for different initial states, aligning with findings from many-body mobility edge studies \cite{yaoManybodyLocalizationBoseHubbard2020,sierantManybodyLocalizationBosons2018}. These results further suggest that the DM approach presented here could serve as a useful tool for detecting phase transitions in experimental regimes where conventional diagnostics are limited.

\section{\label{sec:conclusions} Conclusion}
In this work, we have used an unsupervised learning algorithm based on diffusion maps to learn quantum many-body phase transitions in Bose-Hubbard type models directly from raw measurement configurations. Our results demonstrate that this approach captures multiple equilibrium and dynamical phases in bosonic lattice systems. In particular, the DM method distinguishes phase structure associated with interacting SPT phases hosted in the BHM, as well as ergodic and many-body localized regimes under various settings. Diagnosing these phases conventionally requires accessing quantities that are challenging for experiments, such as nonlocal string correlations and entanglement spectra, or designing probing protocols that require careful control and initial state preparation. In comparison, the unsupervised learning approach presented here can directly operate on experimental measurement snapshots, such as those accessible in current quantum gas microscope platforms \cite{bakr2009quantum,Gross2021ARCM, suTopologicalPhaseTransitions2025}, without requiring the construction of specific diagnostic quantities or tailored data processing.

In addition, while the explicit demonstrations shown here are  for relatively small system sizes and one-dimensional geometry due to the use of  numerical simulations for generating the synthetic measurement data, the diffusion-map workflow itself does not rely on solving the underlying quantum problem and can be applied to experimental datasets at larger system sizes and higher spatial dimensions beyond the reach of these numerical simulations. The algorithm remains computationally efficient for scanning across model parameters in the phase diagram, as it processes data independently at each parameter point rather than requiring a global embedding over aggregated datasets. The models and parameter regimes studied here are directly relevant to ongoing quantum simulator experiments in which the study of quantum phase transitions is an active research direction \cite{suTopologicalPhaseTransitions2025,zivUnsupervisedMachineLearning2025}. Our results therefore provide a practical data-driven framework for exploring quantum many-body phases on such platforms.

\begin{acknowledgments}
We thank the Kavli Institute for Theoretical Physics (KITP) for hospitality while part of this work was completed. We acknowledge support from  the
Air Force Office of Scientific Research (AFOSR) under
Award No. FA9550-24-1-0256.  We also acknowledge the OU Supercomputing Center for Education \& Research (OSCER) at the University of Oklahoma (OU), on which part of the computing for this work was performed.
\end{acknowledgments}

\appendix

\section{Learning many-body localization} \label{sec:appendix}
In the case of learning MBL and thermal phases, the dataset can contain large density variations in the samples space, because the sample configurations tend to localize around the initial state in the MBL phase be broadly distributed in the thermal phase. For such datasets, a common technique implemented in manifold learning is to use density normalization in the kernel \cite{coifmanDiffusionMaps2006,nadlerDiffusionMapsSpectral2006}. Adopting this standard practice, in the application for MBL, we use the kernel $K'_{ij}=K_{ij}/q_iq_j$, with $q_i=\sum_k K_{ik}$ in place of $K_{ij}$, which adds an extra normalization to the Gaussian kernel. 
\begin{figure}[h]
\centering
\includegraphics[width=0.4\textwidth]{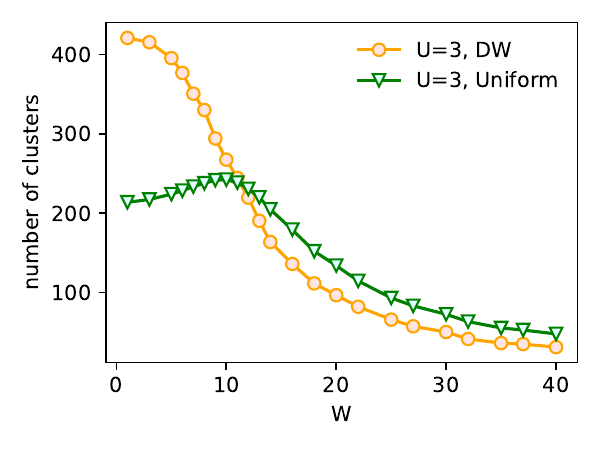}
\caption{Dynamical phases in a disordered chain: comparison between different initial states. Results colored in orange (green) show the DM prediction for an initial state with density-wave (uniform density) pattern, with identical model parameters. }\label{fig:mbl-extra}
\end{figure}

To complement the analysis on the DM's capability for learning MBL in Bose Hubbard systems, here we compare the results obtained from different initial states (uniform and DW) at a strong interaction strength in Fig. \ref{fig:mbl-extra}. In both cases, the results are averaged over 500 disordered realizations. Unlike the case with a weak interaction strength $U=1$ (Fig. \ref{fig:mbl}), where the results for these two initial states are similar, here, a more pronounced difference is observed. For the uniform initial state, which lies lower in the energy spectrum compared to the DW state, the localization also appears to occcur at a higher value of $W$.  This kind of behavior was also reported from analysis on level spacings and transport properties using exact diagonalization for similar parameter regimes \cite{yaoManybodyLocalizationBoseHubbard2020}.

\section{Comparison with PCA analysis}\label{app:pca}
\begin{figure}[h]
\centering
\includegraphics[width=0.4\textwidth]{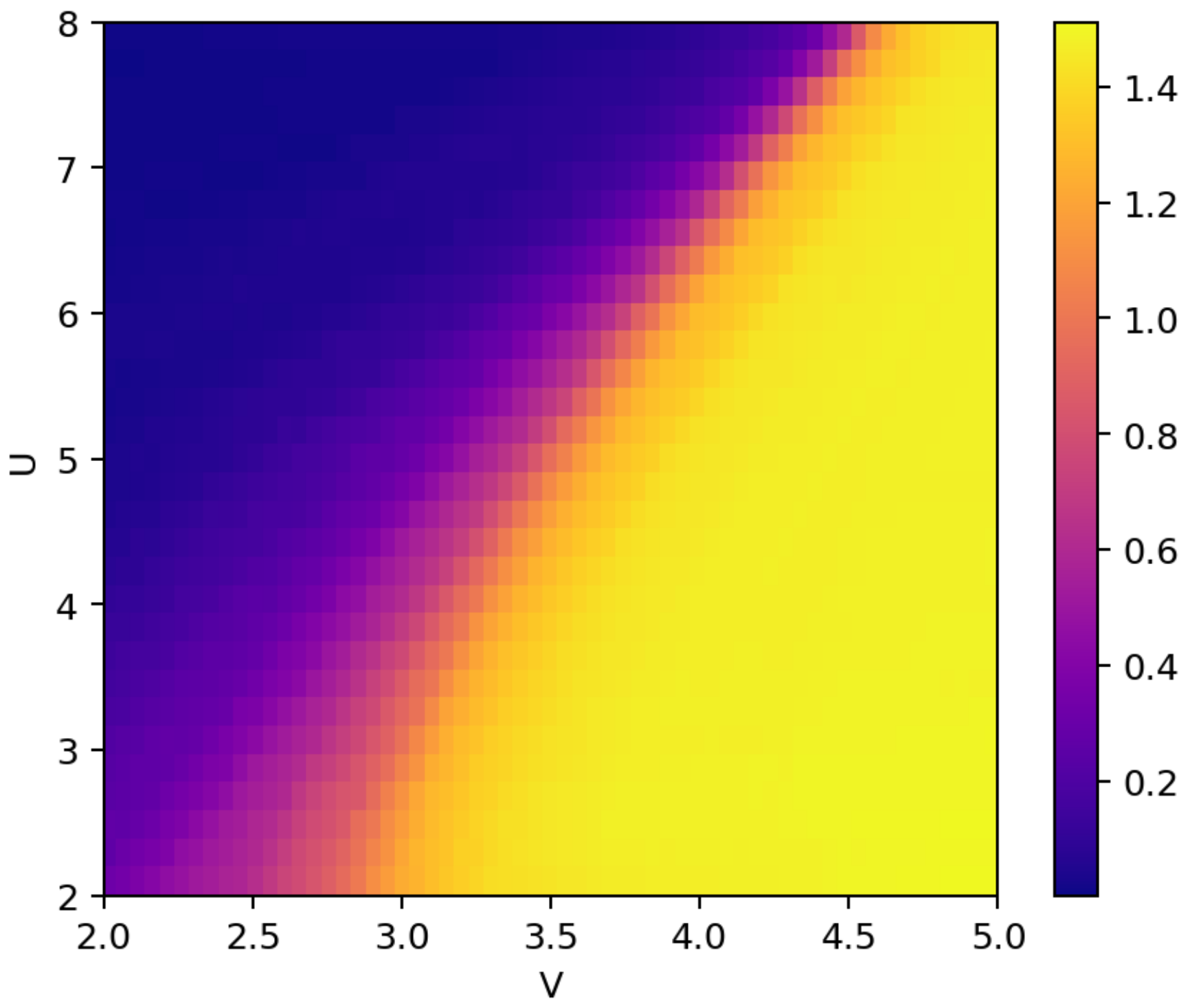}
\caption{Phase diagram obtained from the PCA analysis for the model in Sec. \ref{sec:EBHM} of the main text. The colorbar indicates the cluster label. }\label{fig:pca}
\end{figure}

PCA is a standard dimensionality-reduction method commonly used in unsupervised learning. Here we present the result obtained using PCA for the same measurement data used in Fig.~\ref{fig:EBHM}, providing a baseline comparison with the DM approach. We project the data samples onto the first two principal components and then apply a $k$-means clustering algorithm with $k=3$ to group the samples. We note that, in contrast to this PCA+$k$-means procedure, the DM approach used in the main text does not require the number of phase regions to be specified in advance. As shown in Fig.~\ref{fig:pca}, this PCA-based analysis 
fails to capture the expected phase diagram for this model. This highlights the usefulness of the DM approach for detecting complex phase structure from measurement data.

\bibliography{references}

\begin{thebibliography}{85}%
\makeatletter
\providecommand \@ifxundefined [1]{%
 \@ifx{#1\undefined}
}%
\providecommand \@ifnum [1]{%
 \ifnum #1\expandafter \@firstoftwo
 \else \expandafter \@secondoftwo
 \fi
}%
\providecommand \@ifx [1]{%
 \ifx #1\expandafter \@firstoftwo
 \else \expandafter \@secondoftwo
 \fi
}%
\providecommand \natexlab [1]{#1}%
\providecommand \enquote  [1]{``#1''}%
\providecommand \bibnamefont  [1]{#1}%
\providecommand \bibfnamefont [1]{#1}%
\providecommand \citenamefont [1]{#1}%
\providecommand \href@noop [0]{\@secondoftwo}%
\providecommand \href [0]{\begingroup \@sanitize@url \@href}%
\providecommand \@href[1]{\@@startlink{#1}\@@href}%
\providecommand \@@href[1]{\endgroup#1\@@endlink}%
\providecommand \@sanitize@url [0]{\catcode `\\12\catcode `\$12\catcode `\&12\catcode `\#12\catcode `\^12\catcode `\_12\catcode `\%12\relax}%
\providecommand \@@startlink[1]{}%
\providecommand \@@endlink[0]{}%
\providecommand \url  [0]{\begingroup\@sanitize@url \@url }%
\providecommand \@url [1]{\endgroup\@href {#1}{\urlprefix }}%
\providecommand \urlprefix  [0]{URL }%
\providecommand \Eprint [0]{\href }%
\providecommand \doibase [0]{https://doi.org/}%
\providecommand \selectlanguage [0]{\@gobble}%
\providecommand \bibinfo  [0]{\@secondoftwo}%
\providecommand \bibfield  [0]{\@secondoftwo}%
\providecommand \translation [1]{[#1]}%
\providecommand \BibitemOpen [0]{}%
\providecommand \bibitemStop [0]{}%
\providecommand \bibitemNoStop [0]{.\EOS\space}%
\providecommand \EOS [0]{\spacefactor3000\relax}%
\providecommand \BibitemShut  [1]{\csname bibitem#1\endcsname}%
\let\auto@bib@innerbib\@empty
\bibitem [{\citenamefont {Gross}\ and\ \citenamefont {Bloch}(2017)}]{Gross2017Science}%
  \BibitemOpen
  \bibfield  {author} {\bibinfo {author} {\bibfnamefont {C.}~\bibnamefont {Gross}}\ and\ \bibinfo {author} {\bibfnamefont {I.}~\bibnamefont {Bloch}},\ }\bibfield  {title} {\bibinfo {title} {Quantum simulations with ultracold atoms in optical lattices},\ }\href@noop {} {\bibfield  {journal} {\bibinfo  {journal} {Science}\ }\textbf {\bibinfo {volume} {357}},\ \bibinfo {pages} {995} (\bibinfo {year} {2017})}\BibitemShut {NoStop}%
\bibitem [{\citenamefont {Altman}\ \emph {et~al.}(2021)\citenamefont {Altman}, \citenamefont {Brown}, \citenamefont {Carleo}, \citenamefont {Carr}, \citenamefont {Demler}, \citenamefont {Chin}, \citenamefont {DeMarco}, \citenamefont {Economou}, \citenamefont {Eriksson}, \citenamefont {Fu}, \citenamefont {Greiner}, \citenamefont {Hazzard}, \citenamefont {Hulet}, \citenamefont {Koll\'ar}, \citenamefont {Lev}, \citenamefont {Lukin}, \citenamefont {Ma}, \citenamefont {Mi}, \citenamefont {Misra}, \citenamefont {Monroe}, \citenamefont {Murch}, \citenamefont {Nazario}, \citenamefont {Ni}, \citenamefont {Potter}, \citenamefont {Roushan}, \citenamefont {Saffman}, \citenamefont {Schleier-Smith}, \citenamefont {Siddiqi}, \citenamefont {Simmonds}, \citenamefont {Singh}, \citenamefont {Spielman}, \citenamefont {Temme}, \citenamefont {Weiss}, \citenamefont {Vu\ifmmode \check{c}\else \v{c}\fi{}kovi\ifmmode~\acute{c}\else \'{c}\fi{}}, \citenamefont {Vuleti\ifmmode~\acute{c}\else \'{c}\fi{}}, \citenamefont {Ye},\ and\
  \citenamefont {Zwierlein}}]{Vuletic2021PRXQuantum}%
  \BibitemOpen
  \bibfield  {author} {\bibinfo {author} {\bibfnamefont {E.}~\bibnamefont {Altman}}, \bibinfo {author} {\bibfnamefont {K.~R.}\ \bibnamefont {Brown}}, \bibinfo {author} {\bibfnamefont {G.}~\bibnamefont {Carleo}}, \bibinfo {author} {\bibfnamefont {L.~D.}\ \bibnamefont {Carr}}, \bibinfo {author} {\bibfnamefont {E.}~\bibnamefont {Demler}}, \bibinfo {author} {\bibfnamefont {C.}~\bibnamefont {Chin}}, \bibinfo {author} {\bibfnamefont {B.}~\bibnamefont {DeMarco}}, \bibinfo {author} {\bibfnamefont {S.~E.}\ \bibnamefont {Economou}}, \bibinfo {author} {\bibfnamefont {M.~A.}\ \bibnamefont {Eriksson}}, \bibinfo {author} {\bibfnamefont {K.-M.~C.}\ \bibnamefont {Fu}}, \bibinfo {author} {\bibfnamefont {M.}~\bibnamefont {Greiner}}, \bibinfo {author} {\bibfnamefont {K.~R.}\ \bibnamefont {Hazzard}}, \bibinfo {author} {\bibfnamefont {R.~G.}\ \bibnamefont {Hulet}}, \bibinfo {author} {\bibfnamefont {A.~J.}\ \bibnamefont {Koll\'ar}}, \bibinfo {author} {\bibfnamefont {B.~L.}\ \bibnamefont {Lev}}, \bibinfo {author} {\bibfnamefont
  {M.~D.}\ \bibnamefont {Lukin}}, \bibinfo {author} {\bibfnamefont {R.}~\bibnamefont {Ma}}, \bibinfo {author} {\bibfnamefont {X.}~\bibnamefont {Mi}}, \bibinfo {author} {\bibfnamefont {S.}~\bibnamefont {Misra}}, \bibinfo {author} {\bibfnamefont {C.}~\bibnamefont {Monroe}}, \bibinfo {author} {\bibfnamefont {K.}~\bibnamefont {Murch}}, \bibinfo {author} {\bibfnamefont {Z.}~\bibnamefont {Nazario}}, \bibinfo {author} {\bibfnamefont {K.-K.}\ \bibnamefont {Ni}}, \bibinfo {author} {\bibfnamefont {A.~C.}\ \bibnamefont {Potter}}, \bibinfo {author} {\bibfnamefont {P.}~\bibnamefont {Roushan}}, \bibinfo {author} {\bibfnamefont {M.}~\bibnamefont {Saffman}}, \bibinfo {author} {\bibfnamefont {M.}~\bibnamefont {Schleier-Smith}}, \bibinfo {author} {\bibfnamefont {I.}~\bibnamefont {Siddiqi}}, \bibinfo {author} {\bibfnamefont {R.}~\bibnamefont {Simmonds}}, \bibinfo {author} {\bibfnamefont {M.}~\bibnamefont {Singh}}, \bibinfo {author} {\bibfnamefont {I.}~\bibnamefont {Spielman}}, \bibinfo {author} {\bibfnamefont {K.}~\bibnamefont
  {Temme}}, \bibinfo {author} {\bibfnamefont {D.~S.}\ \bibnamefont {Weiss}}, \bibinfo {author} {\bibfnamefont {J.}~\bibnamefont {Vu\ifmmode \check{c}\else \v{c}\fi{}kovi\ifmmode~\acute{c}\else \'{c}\fi{}}}, \bibinfo {author} {\bibfnamefont {V.}~\bibnamefont {Vuleti\ifmmode~\acute{c}\else \'{c}\fi{}}}, \bibinfo {author} {\bibfnamefont {J.}~\bibnamefont {Ye}},\ and\ \bibinfo {author} {\bibfnamefont {M.}~\bibnamefont {Zwierlein}},\ }\bibfield  {title} {\bibinfo {title} {Quantum simulators: Architectures and opportunities},\ }\href {https://doi.org/10.1103/PRXQuantum.2.017003} {\bibfield  {journal} {\bibinfo  {journal} {PRX Quantum}\ }\textbf {\bibinfo {volume} {2}},\ \bibinfo {pages} {017003} (\bibinfo {year} {2021})}\BibitemShut {NoStop}%
\bibitem [{\citenamefont {Bernien}\ \emph {et~al.}(2017)\citenamefont {Bernien}, \citenamefont {Schwartz}, \citenamefont {Keesling}, \citenamefont {Levine}, \citenamefont {Omran}, \citenamefont {Pichler}, \citenamefont {Choi}, \citenamefont {Zibrov}, \citenamefont {Endres}, \citenamefont {Greiner} \emph {et~al.}}]{Bernien2017Nature51Atom}%
  \BibitemOpen
  \bibfield  {author} {\bibinfo {author} {\bibfnamefont {H.}~\bibnamefont {Bernien}}, \bibinfo {author} {\bibfnamefont {S.}~\bibnamefont {Schwartz}}, \bibinfo {author} {\bibfnamefont {A.}~\bibnamefont {Keesling}}, \bibinfo {author} {\bibfnamefont {H.}~\bibnamefont {Levine}}, \bibinfo {author} {\bibfnamefont {A.}~\bibnamefont {Omran}}, \bibinfo {author} {\bibfnamefont {H.}~\bibnamefont {Pichler}}, \bibinfo {author} {\bibfnamefont {S.}~\bibnamefont {Choi}}, \bibinfo {author} {\bibfnamefont {A.~S.}\ \bibnamefont {Zibrov}}, \bibinfo {author} {\bibfnamefont {M.}~\bibnamefont {Endres}}, \bibinfo {author} {\bibfnamefont {M.}~\bibnamefont {Greiner}}, \emph {et~al.},\ }\bibfield  {title} {\bibinfo {title} {Probing many-body dynamics on a 51-atom quantum simulator},\ }\href@noop {} {\bibfield  {journal} {\bibinfo  {journal} {Nature}\ }\textbf {\bibinfo {volume} {551}},\ \bibinfo {pages} {579} (\bibinfo {year} {2017})}\BibitemShut {NoStop}%
\bibitem [{\citenamefont {Bluvstein}\ \emph {et~al.}(2021)\citenamefont {Bluvstein}, \citenamefont {Omran}, \citenamefont {Levine}, \citenamefont {Keesling}, \citenamefont {Semeghini}, \citenamefont {Ebadi}, \citenamefont {Wang}, \citenamefont {Michailidis}, \citenamefont {Maskara}, \citenamefont {Ho} \emph {et~al.}}]{Bluvstein2021ScienceDrivenRydberg}%
  \BibitemOpen
  \bibfield  {author} {\bibinfo {author} {\bibfnamefont {D.}~\bibnamefont {Bluvstein}}, \bibinfo {author} {\bibfnamefont {A.}~\bibnamefont {Omran}}, \bibinfo {author} {\bibfnamefont {H.}~\bibnamefont {Levine}}, \bibinfo {author} {\bibfnamefont {A.}~\bibnamefont {Keesling}}, \bibinfo {author} {\bibfnamefont {G.}~\bibnamefont {Semeghini}}, \bibinfo {author} {\bibfnamefont {S.}~\bibnamefont {Ebadi}}, \bibinfo {author} {\bibfnamefont {T.~T.}\ \bibnamefont {Wang}}, \bibinfo {author} {\bibfnamefont {A.~A.}\ \bibnamefont {Michailidis}}, \bibinfo {author} {\bibfnamefont {N.}~\bibnamefont {Maskara}}, \bibinfo {author} {\bibfnamefont {W.~W.}\ \bibnamefont {Ho}}, \emph {et~al.},\ }\bibfield  {title} {\bibinfo {title} {Controlling quantum many-body dynamics in driven rydberg atom arrays},\ }\href@noop {} {\bibfield  {journal} {\bibinfo  {journal} {Science}\ }\textbf {\bibinfo {volume} {371}},\ \bibinfo {pages} {1355} (\bibinfo {year} {2021})}\BibitemShut {NoStop}%
\bibitem [{\citenamefont {Ebadi}\ \emph {et~al.}(2021)\citenamefont {Ebadi}, \citenamefont {Wang}, \citenamefont {Levine}, \citenamefont {Keesling}, \citenamefont {Semeghini}, \citenamefont {Omran}, \citenamefont {Bluvstein}, \citenamefont {Samajdar}, \citenamefont {Pichler}, \citenamefont {Ho} \emph {et~al.}}]{Ebadi2021Nature256Atom}%
  \BibitemOpen
  \bibfield  {author} {\bibinfo {author} {\bibfnamefont {S.}~\bibnamefont {Ebadi}}, \bibinfo {author} {\bibfnamefont {T.~T.}\ \bibnamefont {Wang}}, \bibinfo {author} {\bibfnamefont {H.}~\bibnamefont {Levine}}, \bibinfo {author} {\bibfnamefont {A.}~\bibnamefont {Keesling}}, \bibinfo {author} {\bibfnamefont {G.}~\bibnamefont {Semeghini}}, \bibinfo {author} {\bibfnamefont {A.}~\bibnamefont {Omran}}, \bibinfo {author} {\bibfnamefont {D.}~\bibnamefont {Bluvstein}}, \bibinfo {author} {\bibfnamefont {R.}~\bibnamefont {Samajdar}}, \bibinfo {author} {\bibfnamefont {H.}~\bibnamefont {Pichler}}, \bibinfo {author} {\bibfnamefont {W.~W.}\ \bibnamefont {Ho}}, \emph {et~al.},\ }\bibfield  {title} {\bibinfo {title} {Quantum phases of matter on a 256-atom programmable quantum simulator},\ }\href@noop {} {\bibfield  {journal} {\bibinfo  {journal} {Nature}\ }\textbf {\bibinfo {volume} {595}},\ \bibinfo {pages} {227} (\bibinfo {year} {2021})}\BibitemShut {NoStop}%
\bibitem [{\citenamefont {Semeghini}\ \emph {et~al.}(2021)\citenamefont {Semeghini}, \citenamefont {Levine}, \citenamefont {Keesling}, \citenamefont {Ebadi}, \citenamefont {Wang}, \citenamefont {Bluvstein}, \citenamefont {Verresen}, \citenamefont {Pichler}, \citenamefont {Kalinowski}, \citenamefont {Samajdar}, \citenamefont {Omran}, \citenamefont {Sachdev}, \citenamefont {Vishwanath}, \citenamefont {Greiner}, \citenamefont {Vuletić},\ and\ \citenamefont {Lukin}}]{Semeghini2021ScienceSpinLiquid}%
  \BibitemOpen
  \bibfield  {author} {\bibinfo {author} {\bibfnamefont {G.}~\bibnamefont {Semeghini}}, \bibinfo {author} {\bibfnamefont {H.}~\bibnamefont {Levine}}, \bibinfo {author} {\bibfnamefont {A.}~\bibnamefont {Keesling}}, \bibinfo {author} {\bibfnamefont {S.}~\bibnamefont {Ebadi}}, \bibinfo {author} {\bibfnamefont {T.~T.}\ \bibnamefont {Wang}}, \bibinfo {author} {\bibfnamefont {D.}~\bibnamefont {Bluvstein}}, \bibinfo {author} {\bibfnamefont {R.}~\bibnamefont {Verresen}}, \bibinfo {author} {\bibfnamefont {H.}~\bibnamefont {Pichler}}, \bibinfo {author} {\bibfnamefont {M.}~\bibnamefont {Kalinowski}}, \bibinfo {author} {\bibfnamefont {R.}~\bibnamefont {Samajdar}}, \bibinfo {author} {\bibfnamefont {A.}~\bibnamefont {Omran}}, \bibinfo {author} {\bibfnamefont {S.}~\bibnamefont {Sachdev}}, \bibinfo {author} {\bibfnamefont {A.}~\bibnamefont {Vishwanath}}, \bibinfo {author} {\bibfnamefont {M.}~\bibnamefont {Greiner}}, \bibinfo {author} {\bibfnamefont {V.}~\bibnamefont {Vuletić}},\ and\ \bibinfo {author} {\bibfnamefont
  {M.~D.}\ \bibnamefont {Lukin}},\ }\bibfield  {title} {\bibinfo {title} {Probing topological spin liquids on a programmable quantum simulator},\ }\href {https://doi.org/10.1126/science.abi8794} {\bibfield  {journal} {\bibinfo  {journal} {Science}\ }\textbf {\bibinfo {volume} {374}},\ \bibinfo {pages} {1242} (\bibinfo {year} {2021})}\BibitemShut {NoStop}%
\bibitem [{\citenamefont {Manovitz}\ \emph {et~al.}(2025)\citenamefont {Manovitz}, \citenamefont {Li}, \citenamefont {Ebadi}, \citenamefont {Samajdar}, \citenamefont {Geim}, \citenamefont {Evered}, \citenamefont {Bluvstein}, \citenamefont {Zhou}, \citenamefont {Koyluoglu}, \citenamefont {Feldmeier} \emph {et~al.}}]{Manovitz2025NatureCoarsening}%
  \BibitemOpen
  \bibfield  {author} {\bibinfo {author} {\bibfnamefont {T.}~\bibnamefont {Manovitz}}, \bibinfo {author} {\bibfnamefont {S.~H.}\ \bibnamefont {Li}}, \bibinfo {author} {\bibfnamefont {S.}~\bibnamefont {Ebadi}}, \bibinfo {author} {\bibfnamefont {R.}~\bibnamefont {Samajdar}}, \bibinfo {author} {\bibfnamefont {A.~A.}\ \bibnamefont {Geim}}, \bibinfo {author} {\bibfnamefont {S.~J.}\ \bibnamefont {Evered}}, \bibinfo {author} {\bibfnamefont {D.}~\bibnamefont {Bluvstein}}, \bibinfo {author} {\bibfnamefont {H.}~\bibnamefont {Zhou}}, \bibinfo {author} {\bibfnamefont {N.~U.}\ \bibnamefont {Koyluoglu}}, \bibinfo {author} {\bibfnamefont {J.}~\bibnamefont {Feldmeier}}, \emph {et~al.},\ }\bibfield  {title} {\bibinfo {title} {Quantum coarsening and collective dynamics on a programmable simulator},\ }\href@noop {} {\bibfield  {journal} {\bibinfo  {journal} {Nature}\ }\textbf {\bibinfo {volume} {638}},\ \bibinfo {pages} {86} (\bibinfo {year} {2025})}\BibitemShut {NoStop}%
\bibitem [{\citenamefont {Martinez}\ \emph {et~al.}(2016)\citenamefont {Martinez}, \citenamefont {Muschik}, \citenamefont {Schindler}, \citenamefont {Nigg}, \citenamefont {Erhard}, \citenamefont {Heyl}, \citenamefont {Hauke}, \citenamefont {Dalmonte}, \citenamefont {Monz}, \citenamefont {Zoller} \emph {et~al.}}]{Martinez2016NatureSchwinger}%
  \BibitemOpen
  \bibfield  {author} {\bibinfo {author} {\bibfnamefont {E.~A.}\ \bibnamefont {Martinez}}, \bibinfo {author} {\bibfnamefont {C.~A.}\ \bibnamefont {Muschik}}, \bibinfo {author} {\bibfnamefont {P.}~\bibnamefont {Schindler}}, \bibinfo {author} {\bibfnamefont {D.}~\bibnamefont {Nigg}}, \bibinfo {author} {\bibfnamefont {A.}~\bibnamefont {Erhard}}, \bibinfo {author} {\bibfnamefont {M.}~\bibnamefont {Heyl}}, \bibinfo {author} {\bibfnamefont {P.}~\bibnamefont {Hauke}}, \bibinfo {author} {\bibfnamefont {M.}~\bibnamefont {Dalmonte}}, \bibinfo {author} {\bibfnamefont {T.}~\bibnamefont {Monz}}, \bibinfo {author} {\bibfnamefont {P.}~\bibnamefont {Zoller}}, \emph {et~al.},\ }\bibfield  {title} {\bibinfo {title} {Real-time dynamics of lattice gauge theories with a few-qubit quantum computer},\ }\href@noop {} {\bibfield  {journal} {\bibinfo  {journal} {Nature}\ }\textbf {\bibinfo {volume} {534}},\ \bibinfo {pages} {516} (\bibinfo {year} {2016})}\BibitemShut {NoStop}%
\bibitem [{\citenamefont {Smith}\ \emph {et~al.}(2016)\citenamefont {Smith}, \citenamefont {Lee}, \citenamefont {Richerme}, \citenamefont {Neyenhuis}, \citenamefont {Hess}, \citenamefont {Hauke}, \citenamefont {Heyl}, \citenamefont {Huse},\ and\ \citenamefont {Monroe}}]{Smith2016NatPhysMBL}%
  \BibitemOpen
  \bibfield  {author} {\bibinfo {author} {\bibfnamefont {J.}~\bibnamefont {Smith}}, \bibinfo {author} {\bibfnamefont {A.}~\bibnamefont {Lee}}, \bibinfo {author} {\bibfnamefont {P.}~\bibnamefont {Richerme}}, \bibinfo {author} {\bibfnamefont {B.}~\bibnamefont {Neyenhuis}}, \bibinfo {author} {\bibfnamefont {P.~W.}\ \bibnamefont {Hess}}, \bibinfo {author} {\bibfnamefont {P.}~\bibnamefont {Hauke}}, \bibinfo {author} {\bibfnamefont {M.}~\bibnamefont {Heyl}}, \bibinfo {author} {\bibfnamefont {D.~A.}\ \bibnamefont {Huse}},\ and\ \bibinfo {author} {\bibfnamefont {C.}~\bibnamefont {Monroe}},\ }\bibfield  {title} {\bibinfo {title} {Many-body localization in a quantum simulator with programmable random disorder},\ }\href@noop {} {\bibfield  {journal} {\bibinfo  {journal} {Nature Physics}\ }\textbf {\bibinfo {volume} {12}},\ \bibinfo {pages} {907} (\bibinfo {year} {2016})}\BibitemShut {NoStop}%
\bibitem [{\citenamefont {Zhang}\ \emph {et~al.}(2017)\citenamefont {Zhang}, \citenamefont {Pagano}, \citenamefont {Hess}, \citenamefont {Kyprianidis}, \citenamefont {Becker}, \citenamefont {Kaplan}, \citenamefont {Gorshkov}, \citenamefont {Gong},\ and\ \citenamefont {Monroe}}]{Zhang2017NatureDynamicalPT}%
  \BibitemOpen
  \bibfield  {author} {\bibinfo {author} {\bibfnamefont {J.}~\bibnamefont {Zhang}}, \bibinfo {author} {\bibfnamefont {G.}~\bibnamefont {Pagano}}, \bibinfo {author} {\bibfnamefont {P.~W.}\ \bibnamefont {Hess}}, \bibinfo {author} {\bibfnamefont {A.}~\bibnamefont {Kyprianidis}}, \bibinfo {author} {\bibfnamefont {P.}~\bibnamefont {Becker}}, \bibinfo {author} {\bibfnamefont {H.}~\bibnamefont {Kaplan}}, \bibinfo {author} {\bibfnamefont {A.~V.}\ \bibnamefont {Gorshkov}}, \bibinfo {author} {\bibfnamefont {Z.-X.}\ \bibnamefont {Gong}},\ and\ \bibinfo {author} {\bibfnamefont {C.}~\bibnamefont {Monroe}},\ }\bibfield  {title} {\bibinfo {title} {Observation of a many-body dynamical phase transition with a 53-qubit quantum simulator},\ }\href@noop {} {\bibfield  {journal} {\bibinfo  {journal} {Nature}\ }\textbf {\bibinfo {volume} {551}},\ \bibinfo {pages} {601} (\bibinfo {year} {2017})}\BibitemShut {NoStop}%
\bibitem [{\citenamefont {Satzinger}\ \emph {et~al.}(2021)\citenamefont {Satzinger}, \citenamefont {Liu}, \citenamefont {Smith}, \citenamefont {Knapp}, \citenamefont {Newman}, \citenamefont {Jones}, \citenamefont {Chen}, \citenamefont {Quintana}, \citenamefont {Mi}, \citenamefont {Dunsworth} \emph {et~al.}}]{Satzinger2021ScienceToricCode}%
  \BibitemOpen
  \bibfield  {author} {\bibinfo {author} {\bibfnamefont {K.~J.}\ \bibnamefont {Satzinger}}, \bibinfo {author} {\bibfnamefont {Y.-J.}\ \bibnamefont {Liu}}, \bibinfo {author} {\bibfnamefont {A.}~\bibnamefont {Smith}}, \bibinfo {author} {\bibfnamefont {C.}~\bibnamefont {Knapp}}, \bibinfo {author} {\bibfnamefont {M.}~\bibnamefont {Newman}}, \bibinfo {author} {\bibfnamefont {C.}~\bibnamefont {Jones}}, \bibinfo {author} {\bibfnamefont {Z.}~\bibnamefont {Chen}}, \bibinfo {author} {\bibfnamefont {C.}~\bibnamefont {Quintana}}, \bibinfo {author} {\bibfnamefont {X.}~\bibnamefont {Mi}}, \bibinfo {author} {\bibfnamefont {A.}~\bibnamefont {Dunsworth}}, \emph {et~al.},\ }\bibfield  {title} {\bibinfo {title} {Realizing topologically ordered states on a quantum processor},\ }\href@noop {} {\bibfield  {journal} {\bibinfo  {journal} {Science}\ }\textbf {\bibinfo {volume} {374}},\ \bibinfo {pages} {1237} (\bibinfo {year} {2021})}\BibitemShut {NoStop}%
\bibitem [{\citenamefont {Andersen}\ \emph {et~al.}(2025)\citenamefont {Andersen}, \citenamefont {Astrakhantsev}, \citenamefont {Karamlou}, \citenamefont {Berndtsson}, \citenamefont {Motruk}, \citenamefont {Szasz}, \citenamefont {Gross}, \citenamefont {Schuckert}, \citenamefont {Westerhout}, \citenamefont {Zhang} \emph {et~al.}}]{Andersen2025NatureAnalogueDigital}%
  \BibitemOpen
  \bibfield  {author} {\bibinfo {author} {\bibfnamefont {T.~I.}\ \bibnamefont {Andersen}}, \bibinfo {author} {\bibfnamefont {N.}~\bibnamefont {Astrakhantsev}}, \bibinfo {author} {\bibfnamefont {A.~H.}\ \bibnamefont {Karamlou}}, \bibinfo {author} {\bibfnamefont {J.}~\bibnamefont {Berndtsson}}, \bibinfo {author} {\bibfnamefont {J.}~\bibnamefont {Motruk}}, \bibinfo {author} {\bibfnamefont {A.}~\bibnamefont {Szasz}}, \bibinfo {author} {\bibfnamefont {J.~A.}\ \bibnamefont {Gross}}, \bibinfo {author} {\bibfnamefont {A.}~\bibnamefont {Schuckert}}, \bibinfo {author} {\bibfnamefont {T.}~\bibnamefont {Westerhout}}, \bibinfo {author} {\bibfnamefont {Y.}~\bibnamefont {Zhang}}, \emph {et~al.},\ }\bibfield  {title} {\bibinfo {title} {Thermalization and criticality on an analogue--digital quantum simulator},\ }\href@noop {} {\bibfield  {journal} {\bibinfo  {journal} {Nature}\ }\textbf {\bibinfo {volume} {638}},\ \bibinfo {pages} {79} (\bibinfo {year} {2025})}\BibitemShut {NoStop}%
\bibitem [{\citenamefont {Mi}\ \emph {et~al.}(2022)\citenamefont {Mi}, \citenamefont {Ippoliti}, \citenamefont {Quintana}, \citenamefont {Greene}, \citenamefont {Chen}, \citenamefont {Gross}, \citenamefont {Arute}, \citenamefont {Arya}, \citenamefont {Atalaya}, \citenamefont {Babbush} \emph {et~al.}}]{Mi2022NatureTimeCrystallineEigenstate}%
  \BibitemOpen
  \bibfield  {author} {\bibinfo {author} {\bibfnamefont {X.}~\bibnamefont {Mi}}, \bibinfo {author} {\bibfnamefont {M.}~\bibnamefont {Ippoliti}}, \bibinfo {author} {\bibfnamefont {C.}~\bibnamefont {Quintana}}, \bibinfo {author} {\bibfnamefont {A.}~\bibnamefont {Greene}}, \bibinfo {author} {\bibfnamefont {Z.}~\bibnamefont {Chen}}, \bibinfo {author} {\bibfnamefont {J.}~\bibnamefont {Gross}}, \bibinfo {author} {\bibfnamefont {F.}~\bibnamefont {Arute}}, \bibinfo {author} {\bibfnamefont {K.}~\bibnamefont {Arya}}, \bibinfo {author} {\bibfnamefont {J.}~\bibnamefont {Atalaya}}, \bibinfo {author} {\bibfnamefont {R.}~\bibnamefont {Babbush}}, \emph {et~al.},\ }\bibfield  {title} {\bibinfo {title} {Time-crystalline eigenstate order on a quantum processor},\ }\href@noop {} {\bibfield  {journal} {\bibinfo  {journal} {Nature}\ }\textbf {\bibinfo {volume} {601}},\ \bibinfo {pages} {531} (\bibinfo {year} {2022})}\BibitemShut {NoStop}%
\bibitem [{\citenamefont {Lebrat}\ \emph {et~al.}(2024)\citenamefont {Lebrat}, \citenamefont {Xu}, \citenamefont {Kendrick}, \citenamefont {Kale}, \citenamefont {Gang}, \citenamefont {Seetharaman}, \citenamefont {Morera}, \citenamefont {Khatami}, \citenamefont {Demler},\ and\ \citenamefont {Greiner}}]{lebrat2024observation}%
  \BibitemOpen
  \bibfield  {author} {\bibinfo {author} {\bibfnamefont {M.}~\bibnamefont {Lebrat}}, \bibinfo {author} {\bibfnamefont {M.}~\bibnamefont {Xu}}, \bibinfo {author} {\bibfnamefont {L.~H.}\ \bibnamefont {Kendrick}}, \bibinfo {author} {\bibfnamefont {A.}~\bibnamefont {Kale}}, \bibinfo {author} {\bibfnamefont {Y.}~\bibnamefont {Gang}}, \bibinfo {author} {\bibfnamefont {P.}~\bibnamefont {Seetharaman}}, \bibinfo {author} {\bibfnamefont {I.}~\bibnamefont {Morera}}, \bibinfo {author} {\bibfnamefont {E.}~\bibnamefont {Khatami}}, \bibinfo {author} {\bibfnamefont {E.}~\bibnamefont {Demler}},\ and\ \bibinfo {author} {\bibfnamefont {M.}~\bibnamefont {Greiner}},\ }\bibfield  {title} {\bibinfo {title} {Observation of nagaoka polarons in a fermi--hubbard quantum simulator},\ }\href@noop {} {\bibfield  {journal} {\bibinfo  {journal} {Nature}\ }\textbf {\bibinfo {volume} {629}},\ \bibinfo {pages} {317} (\bibinfo {year} {2024})}\BibitemShut {NoStop}%
\bibitem [{\citenamefont {L{\'e}onard}\ \emph {et~al.}(2023)\citenamefont {L{\'e}onard}, \citenamefont {Kim}, \citenamefont {Kwan}, \citenamefont {Segura}, \citenamefont {Grusdt}, \citenamefont {Repellin}, \citenamefont {Goldman},\ and\ \citenamefont {Greiner}}]{leonard2023realization}%
  \BibitemOpen
  \bibfield  {author} {\bibinfo {author} {\bibfnamefont {J.}~\bibnamefont {L{\'e}onard}}, \bibinfo {author} {\bibfnamefont {S.}~\bibnamefont {Kim}}, \bibinfo {author} {\bibfnamefont {J.}~\bibnamefont {Kwan}}, \bibinfo {author} {\bibfnamefont {P.}~\bibnamefont {Segura}}, \bibinfo {author} {\bibfnamefont {F.}~\bibnamefont {Grusdt}}, \bibinfo {author} {\bibfnamefont {C.}~\bibnamefont {Repellin}}, \bibinfo {author} {\bibfnamefont {N.}~\bibnamefont {Goldman}},\ and\ \bibinfo {author} {\bibfnamefont {M.}~\bibnamefont {Greiner}},\ }\bibfield  {title} {\bibinfo {title} {Realization of a fractional quantum hall state with ultracold atoms},\ }\href@noop {} {\bibfield  {journal} {\bibinfo  {journal} {Nature}\ }\textbf {\bibinfo {volume} {619}},\ \bibinfo {pages} {495} (\bibinfo {year} {2023})}\BibitemShut {NoStop}%
\bibitem [{\citenamefont {Xu}\ \emph {et~al.}(2023)\citenamefont {Xu}, \citenamefont {Kendrick}, \citenamefont {Kale}, \citenamefont {Gang}, \citenamefont {Ji}, \citenamefont {Scalettar}, \citenamefont {Lebrat},\ and\ \citenamefont {Greiner}}]{xu2023frustration}%
  \BibitemOpen
  \bibfield  {author} {\bibinfo {author} {\bibfnamefont {M.}~\bibnamefont {Xu}}, \bibinfo {author} {\bibfnamefont {L.~H.}\ \bibnamefont {Kendrick}}, \bibinfo {author} {\bibfnamefont {A.}~\bibnamefont {Kale}}, \bibinfo {author} {\bibfnamefont {Y.}~\bibnamefont {Gang}}, \bibinfo {author} {\bibfnamefont {G.}~\bibnamefont {Ji}}, \bibinfo {author} {\bibfnamefont {R.~T.}\ \bibnamefont {Scalettar}}, \bibinfo {author} {\bibfnamefont {M.}~\bibnamefont {Lebrat}},\ and\ \bibinfo {author} {\bibfnamefont {M.}~\bibnamefont {Greiner}},\ }\bibfield  {title} {\bibinfo {title} {Frustration-and doping-induced magnetism in a fermi--hubbard simulator},\ }\href@noop {} {\bibfield  {journal} {\bibinfo  {journal} {Nature}\ }\textbf {\bibinfo {volume} {620}},\ \bibinfo {pages} {971} (\bibinfo {year} {2023})}\BibitemShut {NoStop}%
\bibitem [{\citenamefont {Fisher}\ \emph {et~al.}(1989)\citenamefont {Fisher}, \citenamefont {Weichman}, \citenamefont {Grinstein},\ and\ \citenamefont {Fisher}}]{Fisher1989PRB}%
  \BibitemOpen
  \bibfield  {author} {\bibinfo {author} {\bibfnamefont {M.~P.~A.}\ \bibnamefont {Fisher}}, \bibinfo {author} {\bibfnamefont {P.~B.}\ \bibnamefont {Weichman}}, \bibinfo {author} {\bibfnamefont {G.}~\bibnamefont {Grinstein}},\ and\ \bibinfo {author} {\bibfnamefont {D.~S.}\ \bibnamefont {Fisher}},\ }\bibfield  {title} {\bibinfo {title} {Boson localization and the superfluid-insulator transition},\ }\href {https://doi.org/10.1103/PhysRevB.40.546} {\bibfield  {journal} {\bibinfo  {journal} {Phys. Rev. B}\ }\textbf {\bibinfo {volume} {40}},\ \bibinfo {pages} {546} (\bibinfo {year} {1989})}\BibitemShut {NoStop}%
\bibitem [{\citenamefont {Jaksch}\ \emph {et~al.}(1998)\citenamefont {Jaksch}, \citenamefont {Bruder}, \citenamefont {Cirac}, \citenamefont {Gardiner},\ and\ \citenamefont {Zoller}}]{Jaksch1998PRL}%
  \BibitemOpen
  \bibfield  {author} {\bibinfo {author} {\bibfnamefont {D.}~\bibnamefont {Jaksch}}, \bibinfo {author} {\bibfnamefont {C.}~\bibnamefont {Bruder}}, \bibinfo {author} {\bibfnamefont {J.~I.}\ \bibnamefont {Cirac}}, \bibinfo {author} {\bibfnamefont {C.~W.}\ \bibnamefont {Gardiner}},\ and\ \bibinfo {author} {\bibfnamefont {P.}~\bibnamefont {Zoller}},\ }\bibfield  {title} {\bibinfo {title} {Cold bosonic atoms in optical lattices},\ }\href {https://doi.org/10.1103/PhysRevLett.81.3108} {\bibfield  {journal} {\bibinfo  {journal} {Phys. Rev. Lett.}\ }\textbf {\bibinfo {volume} {81}},\ \bibinfo {pages} {3108} (\bibinfo {year} {1998})}\BibitemShut {NoStop}%
\bibitem [{\citenamefont {Greiner}\ \emph {et~al.}(2002)\citenamefont {Greiner}, \citenamefont {Mandel}, \citenamefont {Esslinger}, \citenamefont {H{\"a}nsch},\ and\ \citenamefont {Bloch}}]{greiner2002quantum}%
  \BibitemOpen
  \bibfield  {author} {\bibinfo {author} {\bibfnamefont {M.}~\bibnamefont {Greiner}}, \bibinfo {author} {\bibfnamefont {O.}~\bibnamefont {Mandel}}, \bibinfo {author} {\bibfnamefont {T.}~\bibnamefont {Esslinger}}, \bibinfo {author} {\bibfnamefont {T.~W.}\ \bibnamefont {H{\"a}nsch}},\ and\ \bibinfo {author} {\bibfnamefont {I.}~\bibnamefont {Bloch}},\ }\bibfield  {title} {\bibinfo {title} {Quantum phase transition from a superfluid to a mott insulator in a gas of ultracold atoms},\ }\href@noop {} {\bibfield  {journal} {\bibinfo  {journal} {{Nature}}\ }\textbf {\bibinfo {volume} {415}},\ \bibinfo {pages} {39} (\bibinfo {year} {2002})}\BibitemShut {NoStop}%
\bibitem [{\citenamefont {Bloch}\ \emph {et~al.}(2008)\citenamefont {Bloch}, \citenamefont {Dalibard},\ and\ \citenamefont {Zwerger}}]{Bloch2008RMP}%
  \BibitemOpen
  \bibfield  {author} {\bibinfo {author} {\bibfnamefont {I.}~\bibnamefont {Bloch}}, \bibinfo {author} {\bibfnamefont {J.}~\bibnamefont {Dalibard}},\ and\ \bibinfo {author} {\bibfnamefont {W.}~\bibnamefont {Zwerger}},\ }\bibfield  {title} {\bibinfo {title} {Many-body physics with ultracold gases},\ }\href {https://doi.org/10.1103/RevModPhys.80.885} {\bibfield  {journal} {\bibinfo  {journal} {Rev. Mod. Phys.}\ }\textbf {\bibinfo {volume} {80}},\ \bibinfo {pages} {885} (\bibinfo {year} {2008})}\BibitemShut {NoStop}%
\bibitem [{\citenamefont {Bakr}\ \emph {et~al.}(2009)\citenamefont {Bakr}, \citenamefont {Gillen}, \citenamefont {Peng}, \citenamefont {F{\"o}lling},\ and\ \citenamefont {Greiner}}]{bakr2009quantum}%
  \BibitemOpen
  \bibfield  {author} {\bibinfo {author} {\bibfnamefont {W.~S.}\ \bibnamefont {Bakr}}, \bibinfo {author} {\bibfnamefont {J.~I.}\ \bibnamefont {Gillen}}, \bibinfo {author} {\bibfnamefont {A.}~\bibnamefont {Peng}}, \bibinfo {author} {\bibfnamefont {S.}~\bibnamefont {F{\"o}lling}},\ and\ \bibinfo {author} {\bibfnamefont {M.}~\bibnamefont {Greiner}},\ }\bibfield  {title} {\bibinfo {title} {A quantum gas microscope for detecting single atoms in a hubbard-regime optical lattice},\ }\href@noop {} {\bibfield  {journal} {\bibinfo  {journal} {Nature}\ }\textbf {\bibinfo {volume} {462}},\ \bibinfo {pages} {74} (\bibinfo {year} {2009})}\BibitemShut {NoStop}%
\bibitem [{\citenamefont {Sherson}\ \emph {et~al.}(2010)\citenamefont {Sherson}, \citenamefont {Weitenberg}, \citenamefont {Endres}, \citenamefont {Cheneau}, \citenamefont {Bloch},\ and\ \citenamefont {Kuhr}}]{sherson2010single}%
  \BibitemOpen
  \bibfield  {author} {\bibinfo {author} {\bibfnamefont {J.~F.}\ \bibnamefont {Sherson}}, \bibinfo {author} {\bibfnamefont {C.}~\bibnamefont {Weitenberg}}, \bibinfo {author} {\bibfnamefont {M.}~\bibnamefont {Endres}}, \bibinfo {author} {\bibfnamefont {M.}~\bibnamefont {Cheneau}}, \bibinfo {author} {\bibfnamefont {I.}~\bibnamefont {Bloch}},\ and\ \bibinfo {author} {\bibfnamefont {S.}~\bibnamefont {Kuhr}},\ }\bibfield  {title} {\bibinfo {title} {Single-atom-resolved fluorescence imaging of an atomic mott insulator},\ }\href@noop {} {\bibfield  {journal} {\bibinfo  {journal} {Nature}\ }\textbf {\bibinfo {volume} {467}},\ \bibinfo {pages} {68} (\bibinfo {year} {2010})}\BibitemShut {NoStop}%
\bibitem [{\citenamefont {Gross}\ and\ \citenamefont {Bakr}(2021)}]{Gross2021ARCM}%
  \BibitemOpen
  \bibfield  {author} {\bibinfo {author} {\bibfnamefont {C.}~\bibnamefont {Gross}}\ and\ \bibinfo {author} {\bibfnamefont {W.~S.}\ \bibnamefont {Bakr}},\ }\bibfield  {title} {\bibinfo {title} {Quantum gas microscopy for single atom and spin detection},\ }\href@noop {} {\bibfield  {journal} {\bibinfo  {journal} {Nature Physics}\ }\textbf {\bibinfo {volume} {17}},\ \bibinfo {pages} {1316} (\bibinfo {year} {2021})}\BibitemShut {NoStop}%
\bibitem [{\citenamefont {Wei}\ \emph {et~al.}(2022)\citenamefont {Wei}, \citenamefont {Rubio-Abadal}, \citenamefont {Ye}, \citenamefont {Machado}, \citenamefont {Kemp}, \citenamefont {Srakaew}, \citenamefont {Hollerith}, \citenamefont {Rui}, \citenamefont {Gopalakrishnan}, \citenamefont {Yao} \emph {et~al.}}]{Wei2022Science}%
  \BibitemOpen
  \bibfield  {author} {\bibinfo {author} {\bibfnamefont {D.}~\bibnamefont {Wei}}, \bibinfo {author} {\bibfnamefont {A.}~\bibnamefont {Rubio-Abadal}}, \bibinfo {author} {\bibfnamefont {B.}~\bibnamefont {Ye}}, \bibinfo {author} {\bibfnamefont {F.}~\bibnamefont {Machado}}, \bibinfo {author} {\bibfnamefont {J.}~\bibnamefont {Kemp}}, \bibinfo {author} {\bibfnamefont {K.}~\bibnamefont {Srakaew}}, \bibinfo {author} {\bibfnamefont {S.}~\bibnamefont {Hollerith}}, \bibinfo {author} {\bibfnamefont {J.}~\bibnamefont {Rui}}, \bibinfo {author} {\bibfnamefont {S.}~\bibnamefont {Gopalakrishnan}}, \bibinfo {author} {\bibfnamefont {N.~Y.}\ \bibnamefont {Yao}}, \emph {et~al.},\ }\bibfield  {title} {\bibinfo {title} {Quantum gas microscopy of kardar-parisi-zhang superdiffusion},\ }\href@noop {} {\bibfield  {journal} {\bibinfo  {journal} {Science}\ }\textbf {\bibinfo {volume} {376}},\ \bibinfo {pages} {716} (\bibinfo {year} {2022})}\BibitemShut {NoStop}%
\bibitem [{\citenamefont {Koehn}\ \emph {et~al.}(2025)\citenamefont {Koehn}, \citenamefont {Parsonage}, \citenamefont {Duncan}, \citenamefont {Kirton}, \citenamefont {Daley}, \citenamefont {Hilker}, \citenamefont {Haller}, \citenamefont {La~Rooij},\ and\ \citenamefont {Kuhr}}]{QGMBoseGlass2025arXiv}%
  \BibitemOpen
  \bibfield  {author} {\bibinfo {author} {\bibfnamefont {L.}~\bibnamefont {Koehn}}, \bibinfo {author} {\bibfnamefont {C.}~\bibnamefont {Parsonage}}, \bibinfo {author} {\bibfnamefont {C.~W.}\ \bibnamefont {Duncan}}, \bibinfo {author} {\bibfnamefont {P.}~\bibnamefont {Kirton}}, \bibinfo {author} {\bibfnamefont {A.~J.}\ \bibnamefont {Daley}}, \bibinfo {author} {\bibfnamefont {T.}~\bibnamefont {Hilker}}, \bibinfo {author} {\bibfnamefont {E.}~\bibnamefont {Haller}}, \bibinfo {author} {\bibfnamefont {A.}~\bibnamefont {La~Rooij}},\ and\ \bibinfo {author} {\bibfnamefont {S.}~\bibnamefont {Kuhr}},\ }\bibfield  {title} {\bibinfo {title} {Quantum-gas microscopy of the bose-glass phase},\ }\href@noop {} {\bibfield  {journal} {\bibinfo  {journal} {arXiv preprint arXiv:2504.13040}\ } (\bibinfo {year} {2025})}\BibitemShut {NoStop}%
\bibitem [{\citenamefont {Su}\ \emph {et~al.}(2025)\citenamefont {Su}, \citenamefont {Sahay}, \citenamefont {Szurek}, \citenamefont {Douglas}, \citenamefont {Markovic}, \citenamefont {Dag}, \citenamefont {Verresen},\ and\ \citenamefont {Greiner}}]{suTopologicalPhaseTransitions2025}%
  \BibitemOpen
  \bibfield  {author} {\bibinfo {author} {\bibfnamefont {L.}~\bibnamefont {Su}}, \bibinfo {author} {\bibfnamefont {R.}~\bibnamefont {Sahay}}, \bibinfo {author} {\bibfnamefont {M.}~\bibnamefont {Szurek}}, \bibinfo {author} {\bibfnamefont {A.}~\bibnamefont {Douglas}}, \bibinfo {author} {\bibfnamefont {O.}~\bibnamefont {Markovic}}, \bibinfo {author} {\bibfnamefont {C.~B.}\ \bibnamefont {Dag}}, \bibinfo {author} {\bibfnamefont {R.}~\bibnamefont {Verresen}},\ and\ \bibinfo {author} {\bibfnamefont {M.}~\bibnamefont {Greiner}},\ }\href {https://doi.org/10.48550/arXiv.2505.17009} {\bibinfo {title} {Topological {{Phase Transitions}} and {{Mixed State Order}} in a {{Hubbard Quantum Simulator}}}} (\bibinfo {year} {2025}),\ \Eprint {https://arxiv.org/abs/2505.17009} {arXiv:2505.17009 [cond-mat]} \BibitemShut {NoStop}%
\bibitem [{\citenamefont {Ma}\ \emph {et~al.}(2019)\citenamefont {Ma}, \citenamefont {Saxberg}, \citenamefont {Owens}, \citenamefont {Leung}, \citenamefont {Lu}, \citenamefont {Simon},\ and\ \citenamefont {Schuster}}]{Ma2019Nature}%
  \BibitemOpen
  \bibfield  {author} {\bibinfo {author} {\bibfnamefont {R.}~\bibnamefont {Ma}}, \bibinfo {author} {\bibfnamefont {B.}~\bibnamefont {Saxberg}}, \bibinfo {author} {\bibfnamefont {C.}~\bibnamefont {Owens}}, \bibinfo {author} {\bibfnamefont {N.}~\bibnamefont {Leung}}, \bibinfo {author} {\bibfnamefont {Y.}~\bibnamefont {Lu}}, \bibinfo {author} {\bibfnamefont {J.}~\bibnamefont {Simon}},\ and\ \bibinfo {author} {\bibfnamefont {D.~I.}\ \bibnamefont {Schuster}},\ }\bibfield  {title} {\bibinfo {title} {A dissipatively stabilized mott insulator of photons},\ }\href@noop {} {\bibfield  {journal} {\bibinfo  {journal} {Nature}\ }\textbf {\bibinfo {volume} {566}},\ \bibinfo {pages} {51} (\bibinfo {year} {2019})}\BibitemShut {NoStop}%
\bibitem [{\citenamefont {Fitzpatrick}\ \emph {et~al.}(2017)\citenamefont {Fitzpatrick}, \citenamefont {Sundaresan}, \citenamefont {Li}, \citenamefont {Koch},\ and\ \citenamefont {Houck}}]{Fitzpatrick2017PRX}%
  \BibitemOpen
  \bibfield  {author} {\bibinfo {author} {\bibfnamefont {M.}~\bibnamefont {Fitzpatrick}}, \bibinfo {author} {\bibfnamefont {N.~M.}\ \bibnamefont {Sundaresan}}, \bibinfo {author} {\bibfnamefont {A.~C.~Y.}\ \bibnamefont {Li}}, \bibinfo {author} {\bibfnamefont {J.}~\bibnamefont {Koch}},\ and\ \bibinfo {author} {\bibfnamefont {A.~A.}\ \bibnamefont {Houck}},\ }\bibfield  {title} {\bibinfo {title} {Observation of a dissipative phase transition in a one-dimensional circuit qed lattice},\ }\href {https://doi.org/10.1103/PhysRevX.7.011016} {\bibfield  {journal} {\bibinfo  {journal} {Phys. Rev. X}\ }\textbf {\bibinfo {volume} {7}},\ \bibinfo {pages} {011016} (\bibinfo {year} {2017})}\BibitemShut {NoStop}%
\bibitem [{\citenamefont {Zhang}\ \emph {et~al.}(2023)\citenamefont {Zhang}, \citenamefont {Kim}, \citenamefont {Mark}, \citenamefont {Choi},\ and\ \citenamefont {Painter}}]{Zhang2023Science}%
  \BibitemOpen
  \bibfield  {author} {\bibinfo {author} {\bibfnamefont {X.}~\bibnamefont {Zhang}}, \bibinfo {author} {\bibfnamefont {E.}~\bibnamefont {Kim}}, \bibinfo {author} {\bibfnamefont {D.~K.}\ \bibnamefont {Mark}}, \bibinfo {author} {\bibfnamefont {S.}~\bibnamefont {Choi}},\ and\ \bibinfo {author} {\bibfnamefont {O.}~\bibnamefont {Painter}},\ }\bibfield  {title} {\bibinfo {title} {A superconducting quantum simulator based on a photonic-bandgap metamaterial},\ }\href@noop {} {\bibfield  {journal} {\bibinfo  {journal} {Science}\ }\textbf {\bibinfo {volume} {379}},\ \bibinfo {pages} {278} (\bibinfo {year} {2023})}\BibitemShut {NoStop}%
\bibitem [{\citenamefont {Du}\ \emph {et~al.}(2024)\citenamefont {Du}, \citenamefont {Suresh}, \citenamefont {L\'opez}, \citenamefont {Cadiente},\ and\ \citenamefont {Ma}}]{Du2024PRL}%
  \BibitemOpen
  \bibfield  {author} {\bibinfo {author} {\bibfnamefont {B.}~\bibnamefont {Du}}, \bibinfo {author} {\bibfnamefont {R.}~\bibnamefont {Suresh}}, \bibinfo {author} {\bibfnamefont {S.}~\bibnamefont {L\'opez}}, \bibinfo {author} {\bibfnamefont {J.}~\bibnamefont {Cadiente}},\ and\ \bibinfo {author} {\bibfnamefont {R.}~\bibnamefont {Ma}},\ }\bibfield  {title} {\bibinfo {title} {Probing site-resolved current in strongly interacting superconducting circuit lattices},\ }\href {https://doi.org/10.1103/PhysRevLett.133.060601} {\bibfield  {journal} {\bibinfo  {journal} {Phys. Rev. Lett.}\ }\textbf {\bibinfo {volume} {133}},\ \bibinfo {pages} {060601} (\bibinfo {year} {2024})}\BibitemShut {NoStop}%
\bibitem [{\citenamefont {Du}\ \emph {et~al.}(2025)\citenamefont {Du}, \citenamefont {Guo}, \citenamefont {L\'opez},\ and\ \citenamefont {Ma}}]{Du2025PRR}%
  \BibitemOpen
  \bibfield  {author} {\bibinfo {author} {\bibfnamefont {B.}~\bibnamefont {Du}}, \bibinfo {author} {\bibfnamefont {Q.}~\bibnamefont {Guo}}, \bibinfo {author} {\bibfnamefont {S.}~\bibnamefont {L\'opez}},\ and\ \bibinfo {author} {\bibfnamefont {R.}~\bibnamefont {Ma}},\ }\bibfield  {title} {\bibinfo {title} {Tunneling spectroscopy in superconducting circuit lattices},\ }\href {https://doi.org/10.1103/PhysRevResearch.7.L022038} {\bibfield  {journal} {\bibinfo  {journal} {Phys. Rev. Res.}\ }\textbf {\bibinfo {volume} {7}},\ \bibinfo {pages} {L022038} (\bibinfo {year} {2025})}\BibitemShut {NoStop}%
\bibitem [{\citenamefont {Kim}\ \emph {et~al.}(2025)\citenamefont {Kim}, \citenamefont {Kang}, \citenamefont {Segura}, \citenamefont {Li}, \citenamefont {Lake}, \citenamefont {Bakkali-Hassani},\ and\ \citenamefont {Greiner}}]{kim2025multi}%
  \BibitemOpen
  \bibfield  {author} {\bibinfo {author} {\bibfnamefont {S.}~\bibnamefont {Kim}}, \bibinfo {author} {\bibfnamefont {B.}~\bibnamefont {Kang}}, \bibinfo {author} {\bibfnamefont {P.}~\bibnamefont {Segura}}, \bibinfo {author} {\bibfnamefont {Y.}~\bibnamefont {Li}}, \bibinfo {author} {\bibfnamefont {E.}~\bibnamefont {Lake}}, \bibinfo {author} {\bibfnamefont {B.}~\bibnamefont {Bakkali-Hassani}},\ and\ \bibinfo {author} {\bibfnamefont {M.}~\bibnamefont {Greiner}},\ }\bibfield  {title} {\bibinfo {title} {Multi-particle quantum walks in a dipole-conserving bose-hubbard model},\ }\href@noop {} {\bibfield  {journal} {\bibinfo  {journal} {arXiv preprint arXiv:2511.02343}\ } (\bibinfo {year} {2025})}\BibitemShut {NoStop}%
\bibitem [{\citenamefont {Wei}\ \emph {et~al.}(2023)\citenamefont {Wei}, \citenamefont {Adler}, \citenamefont {Srakaew}, \citenamefont {Agrawal}, \citenamefont {Weckesser}, \citenamefont {Bloch},\ and\ \citenamefont {Zeiher}}]{Wei2023PRXBraneParity}%
  \BibitemOpen
  \bibfield  {author} {\bibinfo {author} {\bibfnamefont {D.}~\bibnamefont {Wei}}, \bibinfo {author} {\bibfnamefont {D.}~\bibnamefont {Adler}}, \bibinfo {author} {\bibfnamefont {K.}~\bibnamefont {Srakaew}}, \bibinfo {author} {\bibfnamefont {S.}~\bibnamefont {Agrawal}}, \bibinfo {author} {\bibfnamefont {P.}~\bibnamefont {Weckesser}}, \bibinfo {author} {\bibfnamefont {I.}~\bibnamefont {Bloch}},\ and\ \bibinfo {author} {\bibfnamefont {J.}~\bibnamefont {Zeiher}},\ }\bibfield  {title} {\bibinfo {title} {Observation of brane parity order in programmable optical lattices},\ }\href {https://doi.org/10.1103/PhysRevX.13.021042} {\bibfield  {journal} {\bibinfo  {journal} {Phys. Rev. X}\ }\textbf {\bibinfo {volume} {13}},\ \bibinfo {pages} {021042} (\bibinfo {year} {2023})}\BibitemShut {NoStop}%
\bibitem [{\citenamefont {Rubio-Abadal}\ \emph {et~al.}(2020)\citenamefont {Rubio-Abadal}, \citenamefont {Ippoliti}, \citenamefont {Hollerith}, \citenamefont {Wei}, \citenamefont {Rui}, \citenamefont {Sondhi}, \citenamefont {Khemani}, \citenamefont {Gross},\ and\ \citenamefont {Bloch}}]{blochPrethermalPRX2020}%
  \BibitemOpen
  \bibfield  {author} {\bibinfo {author} {\bibfnamefont {A.}~\bibnamefont {Rubio-Abadal}}, \bibinfo {author} {\bibfnamefont {M.}~\bibnamefont {Ippoliti}}, \bibinfo {author} {\bibfnamefont {S.}~\bibnamefont {Hollerith}}, \bibinfo {author} {\bibfnamefont {D.}~\bibnamefont {Wei}}, \bibinfo {author} {\bibfnamefont {J.}~\bibnamefont {Rui}}, \bibinfo {author} {\bibfnamefont {S.~L.}\ \bibnamefont {Sondhi}}, \bibinfo {author} {\bibfnamefont {V.}~\bibnamefont {Khemani}}, \bibinfo {author} {\bibfnamefont {C.}~\bibnamefont {Gross}},\ and\ \bibinfo {author} {\bibfnamefont {I.}~\bibnamefont {Bloch}},\ }\bibfield  {title} {\bibinfo {title} {Floquet prethermalization in a bose-hubbard system},\ }\href {https://doi.org/10.1103/PhysRevX.10.021044} {\bibfield  {journal} {\bibinfo  {journal} {Phys. Rev. X}\ }\textbf {\bibinfo {volume} {10}},\ \bibinfo {pages} {021044} (\bibinfo {year} {2020})}\BibitemShut {NoStop}%
\bibitem [{\citenamefont {Hasan}\ and\ \citenamefont {Kane}(2010)}]{Hasan2010RMP}%
  \BibitemOpen
  \bibfield  {author} {\bibinfo {author} {\bibfnamefont {M.~Z.}\ \bibnamefont {Hasan}}\ and\ \bibinfo {author} {\bibfnamefont {C.~L.}\ \bibnamefont {Kane}},\ }\bibfield  {title} {\bibinfo {title} {Colloquium: Topological insulators},\ }\href {https://doi.org/10.1103/RevModPhys.82.3045} {\bibfield  {journal} {\bibinfo  {journal} {Rev. Mod. Phys.}\ }\textbf {\bibinfo {volume} {82}},\ \bibinfo {pages} {3045} (\bibinfo {year} {2010})}\BibitemShut {NoStop}%
\bibitem [{\citenamefont {Qi}\ and\ \citenamefont {Zhang}(2011)}]{Qi2011RMP}%
  \BibitemOpen
  \bibfield  {author} {\bibinfo {author} {\bibfnamefont {X.-L.}\ \bibnamefont {Qi}}\ and\ \bibinfo {author} {\bibfnamefont {S.-C.}\ \bibnamefont {Zhang}},\ }\bibfield  {title} {\bibinfo {title} {Topological insulators and superconductors},\ }\href {https://doi.org/10.1103/RevModPhys.83.1057} {\bibfield  {journal} {\bibinfo  {journal} {Rev. Mod. Phys.}\ }\textbf {\bibinfo {volume} {83}},\ \bibinfo {pages} {1057} (\bibinfo {year} {2011})}\BibitemShut {NoStop}%
\bibitem [{\citenamefont {Lukin}\ \emph {et~al.}(2019)\citenamefont {Lukin} \emph {et~al.}}]{Lukin2019Science}%
  \BibitemOpen
  \bibfield  {author} {\bibinfo {author} {\bibfnamefont {A.}~\bibnamefont {Lukin}} \emph {et~al.},\ }\bibfield  {title} {\bibinfo {title} {Probing entanglement in a many-body--localized system},\ }\href {https://doi.org/10.1126/science.aau0818} {\bibfield  {journal} {\bibinfo  {journal} {Science}\ }\textbf {\bibinfo {volume} {364}},\ \bibinfo {pages} {256} (\bibinfo {year} {2019})}\BibitemShut {NoStop}%
\bibitem [{\citenamefont {Kaufman}\ \emph {et~al.}(2016)\citenamefont {Kaufman}, \citenamefont {Tai}, \citenamefont {Lukin}, \citenamefont {Rispoli}, \citenamefont {Schittko}, \citenamefont {Preiss},\ and\ \citenamefont {Greiner}}]{kaufman2016quantum}%
  \BibitemOpen
  \bibfield  {author} {\bibinfo {author} {\bibfnamefont {A.~M.}\ \bibnamefont {Kaufman}}, \bibinfo {author} {\bibfnamefont {M.~E.}\ \bibnamefont {Tai}}, \bibinfo {author} {\bibfnamefont {A.}~\bibnamefont {Lukin}}, \bibinfo {author} {\bibfnamefont {M.}~\bibnamefont {Rispoli}}, \bibinfo {author} {\bibfnamefont {R.}~\bibnamefont {Schittko}}, \bibinfo {author} {\bibfnamefont {P.~M.}\ \bibnamefont {Preiss}},\ and\ \bibinfo {author} {\bibfnamefont {M.}~\bibnamefont {Greiner}},\ }\bibfield  {title} {\bibinfo {title} {Quantum thermalization through entanglement in an isolated many-body system},\ }\href@noop {} {\bibfield  {journal} {\bibinfo  {journal} {Science}\ }\textbf {\bibinfo {volume} {353}},\ \bibinfo {pages} {794} (\bibinfo {year} {2016})}\BibitemShut {NoStop}%
\bibitem [{\citenamefont {Abanin}\ \emph {et~al.}(2019)\citenamefont {Abanin}, \citenamefont {Altman}, \citenamefont {Bloch},\ and\ \citenamefont {Serbyn}}]{abaninColloquiumManybodyLocalization2019}%
  \BibitemOpen
  \bibfield  {author} {\bibinfo {author} {\bibfnamefont {D.~A.}\ \bibnamefont {Abanin}}, \bibinfo {author} {\bibfnamefont {E.}~\bibnamefont {Altman}}, \bibinfo {author} {\bibfnamefont {I.}~\bibnamefont {Bloch}},\ and\ \bibinfo {author} {\bibfnamefont {M.}~\bibnamefont {Serbyn}},\ }\bibfield  {title} {\bibinfo {title} {Colloquium: {{Many-body}} localization, thermalization, and entanglement},\ }\href {https://doi.org/10.1103/RevModPhys.91.021001} {\bibfield  {journal} {\bibinfo  {journal} {Reviews of Modern Physics}\ }\textbf {\bibinfo {volume} {91}},\ \bibinfo {pages} {021001} (\bibinfo {year} {2019})}\BibitemShut {NoStop}%
\bibitem [{\citenamefont {Cian}\ \emph {et~al.}(2021)\citenamefont {Cian}, \citenamefont {Dehghani}, \citenamefont {Elben}, \citenamefont {Vermersch}, \citenamefont {Zhu}, \citenamefont {Barkeshli}, \citenamefont {Zoller},\ and\ \citenamefont {Hafezi}}]{chernnumberHafeziPRL2021}%
  \BibitemOpen
  \bibfield  {author} {\bibinfo {author} {\bibfnamefont {Z.-P.}\ \bibnamefont {Cian}}, \bibinfo {author} {\bibfnamefont {H.}~\bibnamefont {Dehghani}}, \bibinfo {author} {\bibfnamefont {A.}~\bibnamefont {Elben}}, \bibinfo {author} {\bibfnamefont {B.}~\bibnamefont {Vermersch}}, \bibinfo {author} {\bibfnamefont {G.}~\bibnamefont {Zhu}}, \bibinfo {author} {\bibfnamefont {M.}~\bibnamefont {Barkeshli}}, \bibinfo {author} {\bibfnamefont {P.}~\bibnamefont {Zoller}},\ and\ \bibinfo {author} {\bibfnamefont {M.}~\bibnamefont {Hafezi}},\ }\bibfield  {title} {\bibinfo {title} {Many-body chern number from statistical correlations of randomized measurements},\ }\href {https://doi.org/10.1103/PhysRevLett.126.050501} {\bibfield  {journal} {\bibinfo  {journal} {Phys. Rev. Lett.}\ }\textbf {\bibinfo {volume} {126}},\ \bibinfo {pages} {050501} (\bibinfo {year} {2021})}\BibitemShut {NoStop}%
\bibitem [{\citenamefont {Elben}\ \emph {et~al.}(2020)\citenamefont {Elben}, \citenamefont {Yu}, \citenamefont {Zhu}, \citenamefont {Hafezi}, \citenamefont {Pollmann}, \citenamefont {Zoller},\ and\ \citenamefont {Vermersch}}]{elben2020many}%
  \BibitemOpen
  \bibfield  {author} {\bibinfo {author} {\bibfnamefont {A.}~\bibnamefont {Elben}}, \bibinfo {author} {\bibfnamefont {J.}~\bibnamefont {Yu}}, \bibinfo {author} {\bibfnamefont {G.}~\bibnamefont {Zhu}}, \bibinfo {author} {\bibfnamefont {M.}~\bibnamefont {Hafezi}}, \bibinfo {author} {\bibfnamefont {F.}~\bibnamefont {Pollmann}}, \bibinfo {author} {\bibfnamefont {P.}~\bibnamefont {Zoller}},\ and\ \bibinfo {author} {\bibfnamefont {B.}~\bibnamefont {Vermersch}},\ }\bibfield  {title} {\bibinfo {title} {Many-body topological invariants from randomized measurements in synthetic quantum matter},\ }\href@noop {} {\bibfield  {journal} {\bibinfo  {journal} {Science advances}\ }\textbf {\bibinfo {volume} {6}},\ \bibinfo {pages} {eaaz3666} (\bibinfo {year} {2020})}\BibitemShut {NoStop}%
\bibitem [{\citenamefont {Carrasquilla}\ and\ \citenamefont {Melko}(2017)}]{Carrasquilla2017NatPhys}%
  \BibitemOpen
  \bibfield  {author} {\bibinfo {author} {\bibfnamefont {J.}~\bibnamefont {Carrasquilla}}\ and\ \bibinfo {author} {\bibfnamefont {R.~G.}\ \bibnamefont {Melko}},\ }\bibfield  {title} {\bibinfo {title} {Machine learning phases of matter},\ }\href@noop {} {\bibfield  {journal} {\bibinfo  {journal} {Nature Physics}\ }\textbf {\bibinfo {volume} {13}},\ \bibinfo {pages} {431} (\bibinfo {year} {2017})}\BibitemShut {NoStop}%
\bibitem [{\citenamefont {Van~Nieuwenburg}\ \emph {et~al.}(2017)\citenamefont {Van~Nieuwenburg}, \citenamefont {Liu},\ and\ \citenamefont {Huber}}]{vanNieuwenburg2017NatPhys}%
  \BibitemOpen
  \bibfield  {author} {\bibinfo {author} {\bibfnamefont {E.~P.}\ \bibnamefont {Van~Nieuwenburg}}, \bibinfo {author} {\bibfnamefont {Y.-H.}\ \bibnamefont {Liu}},\ and\ \bibinfo {author} {\bibfnamefont {S.~D.}\ \bibnamefont {Huber}},\ }\bibfield  {title} {\bibinfo {title} {Learning phase transitions by confusion},\ }\href@noop {} {\bibfield  {journal} {\bibinfo  {journal} {Nature Physics}\ }\textbf {\bibinfo {volume} {13}},\ \bibinfo {pages} {435} (\bibinfo {year} {2017})}\BibitemShut {NoStop}%
\bibitem [{\citenamefont {Rem}\ \emph {et~al.}(2019)\citenamefont {Rem}, \citenamefont {K{\"a}ming}, \citenamefont {Tarnowski}, \citenamefont {Asteria}, \citenamefont {Fl{\"a}schner}, \citenamefont {Becker}, \citenamefont {Sengstock},\ and\ \citenamefont {Weitenberg}}]{phaseNNexpNatPhys2019}%
  \BibitemOpen
  \bibfield  {author} {\bibinfo {author} {\bibfnamefont {B.~S.}\ \bibnamefont {Rem}}, \bibinfo {author} {\bibfnamefont {N.}~\bibnamefont {K{\"a}ming}}, \bibinfo {author} {\bibfnamefont {M.}~\bibnamefont {Tarnowski}}, \bibinfo {author} {\bibfnamefont {L.}~\bibnamefont {Asteria}}, \bibinfo {author} {\bibfnamefont {N.}~\bibnamefont {Fl{\"a}schner}}, \bibinfo {author} {\bibfnamefont {C.}~\bibnamefont {Becker}}, \bibinfo {author} {\bibfnamefont {K.}~\bibnamefont {Sengstock}},\ and\ \bibinfo {author} {\bibfnamefont {C.}~\bibnamefont {Weitenberg}},\ }\bibfield  {title} {\bibinfo {title} {Identifying quantum phase transitions using artificial neural networks on experimental data},\ }\href@noop {} {\bibfield  {journal} {\bibinfo  {journal} {Nature Physics}\ }\textbf {\bibinfo {volume} {15}},\ \bibinfo {pages} {917} (\bibinfo {year} {2019})}\BibitemShut {NoStop}%
\bibitem [{\citenamefont {Wang}(2016)}]{Wang2016PRB}%
  \BibitemOpen
  \bibfield  {author} {\bibinfo {author} {\bibfnamefont {L.}~\bibnamefont {Wang}},\ }\bibfield  {title} {\bibinfo {title} {Discovering phase transitions with unsupervised learning},\ }\href {https://doi.org/10.1103/PhysRevB.94.195105} {\bibfield  {journal} {\bibinfo  {journal} {Phys. Rev. B}\ }\textbf {\bibinfo {volume} {94}},\ \bibinfo {pages} {195105} (\bibinfo {year} {2016})}\BibitemShut {NoStop}%
\bibitem [{\citenamefont {Hu}\ \emph {et~al.}(2017)\citenamefont {Hu}, \citenamefont {Singh},\ and\ \citenamefont {Scalettar}}]{Hu2017PRE}%
  \BibitemOpen
  \bibfield  {author} {\bibinfo {author} {\bibfnamefont {W.}~\bibnamefont {Hu}}, \bibinfo {author} {\bibfnamefont {R.~R.~P.}\ \bibnamefont {Singh}},\ and\ \bibinfo {author} {\bibfnamefont {R.~T.}\ \bibnamefont {Scalettar}},\ }\bibfield  {title} {\bibinfo {title} {Discovering phases, phase transitions, and crossovers through unsupervised machine learning: A critical examination},\ }\href {https://doi.org/10.1103/PhysRevE.95.062122} {\bibfield  {journal} {\bibinfo  {journal} {Phys. Rev. E}\ }\textbf {\bibinfo {volume} {95}},\ \bibinfo {pages} {062122} (\bibinfo {year} {2017})}\BibitemShut {NoStop}%
\bibitem [{\citenamefont {Wetzel}(2017)}]{classicalAE2017}%
  \BibitemOpen
  \bibfield  {author} {\bibinfo {author} {\bibfnamefont {S.~J.}\ \bibnamefont {Wetzel}},\ }\bibfield  {title} {\bibinfo {title} {Unsupervised learning of phase transitions: From principal component analysis to variational autoencoders},\ }\href {https://doi.org/10.1103/PhysRevE.96.022140} {\bibfield  {journal} {\bibinfo  {journal} {Phys. Rev. E}\ }\textbf {\bibinfo {volume} {96}},\ \bibinfo {pages} {022140} (\bibinfo {year} {2017})}\BibitemShut {NoStop}%
\bibitem [{\citenamefont {K{\"a}ming}\ \emph {et~al.}(2021)\citenamefont {K{\"a}ming}, \citenamefont {Dawid}, \citenamefont {Kottmann}, \citenamefont {Lewenstein}, \citenamefont {Sengstock}, \citenamefont {Dauphin},\ and\ \citenamefont {Weitenberg}}]{AEexp2021}%
  \BibitemOpen
  \bibfield  {author} {\bibinfo {author} {\bibfnamefont {N.}~\bibnamefont {K{\"a}ming}}, \bibinfo {author} {\bibfnamefont {A.}~\bibnamefont {Dawid}}, \bibinfo {author} {\bibfnamefont {K.}~\bibnamefont {Kottmann}}, \bibinfo {author} {\bibfnamefont {M.}~\bibnamefont {Lewenstein}}, \bibinfo {author} {\bibfnamefont {K.}~\bibnamefont {Sengstock}}, \bibinfo {author} {\bibfnamefont {A.}~\bibnamefont {Dauphin}},\ and\ \bibinfo {author} {\bibfnamefont {C.}~\bibnamefont {Weitenberg}},\ }\bibfield  {title} {\bibinfo {title} {Unsupervised machine learning of topological phase transitions from experimental data},\ }\href@noop {} {\bibfield  {journal} {\bibinfo  {journal} {Machine Learning: Science and Technology}\ }\textbf {\bibinfo {volume} {2}},\ \bibinfo {pages} {035037} (\bibinfo {year} {2021})}\BibitemShut {NoStop}%
\bibitem [{\citenamefont {Ch'ng}\ \emph {et~al.}(2018)\citenamefont {Ch'ng}, \citenamefont {Vazquez},\ and\ \citenamefont {Khatami}}]{chngUnsupervisedMachineLearning2018a}%
  \BibitemOpen
  \bibfield  {author} {\bibinfo {author} {\bibfnamefont {K.}~\bibnamefont {Ch'ng}}, \bibinfo {author} {\bibfnamefont {N.}~\bibnamefont {Vazquez}},\ and\ \bibinfo {author} {\bibfnamefont {E.}~\bibnamefont {Khatami}},\ }\bibfield  {title} {\bibinfo {title} {Unsupervised machine learning account of magnetic transitions in the {{Hubbard}} model},\ }\href {https://doi.org/10.1103/PhysRevE.97.013306} {\bibfield  {journal} {\bibinfo  {journal} {Physical Review E}\ }\textbf {\bibinfo {volume} {97}},\ \bibinfo {pages} {013306} (\bibinfo {year} {2018})}\BibitemShut {NoStop}%
\bibitem [{\citenamefont {{Rodriguez-Nieva}}\ and\ \citenamefont {Scheurer}(2019)}]{rodriguez-nievaIdentifyingTopologicalOrder2019}%
  \BibitemOpen
  \bibfield  {author} {\bibinfo {author} {\bibfnamefont {J.~F.}\ \bibnamefont {{Rodriguez-Nieva}}}\ and\ \bibinfo {author} {\bibfnamefont {M.~S.}\ \bibnamefont {Scheurer}},\ }\bibfield  {title} {\bibinfo {title} {Identifying topological order through unsupervised machine learning},\ }\href {https://doi.org/10.1038/s41567-019-0512-x} {\bibfield  {journal} {\bibinfo  {journal} {Nature Physics}\ }\textbf {\bibinfo {volume} {15}},\ \bibinfo {pages} {790} (\bibinfo {year} {2019})}\BibitemShut {NoStop}%
\bibitem [{\citenamefont {Lidiak}\ and\ \citenamefont {Gong}(2020)}]{lidiakUnsupervisedMachineLearning2020}%
  \BibitemOpen
  \bibfield  {author} {\bibinfo {author} {\bibfnamefont {A.}~\bibnamefont {Lidiak}}\ and\ \bibinfo {author} {\bibfnamefont {Z.}~\bibnamefont {Gong}},\ }\bibfield  {title} {\bibinfo {title} {Unsupervised {{Machine Learning}} of {{Quantum Phase Transitions Using Diffusion Maps}}},\ }\href {https://doi.org/10.1103/PhysRevLett.125.225701} {\bibfield  {journal} {\bibinfo  {journal} {Physical Review Letters}\ }\textbf {\bibinfo {volume} {125}},\ \bibinfo {pages} {225701} (\bibinfo {year} {2020})}\BibitemShut {NoStop}%
\bibitem [{\citenamefont {Che}\ \emph {et~al.}(2020)\citenamefont {Che}, \citenamefont {Gneiting}, \citenamefont {Liu},\ and\ \citenamefont {Nori}}]{cheTopologicalQuantumPhase2020a}%
  \BibitemOpen
  \bibfield  {author} {\bibinfo {author} {\bibfnamefont {Y.}~\bibnamefont {Che}}, \bibinfo {author} {\bibfnamefont {C.}~\bibnamefont {Gneiting}}, \bibinfo {author} {\bibfnamefont {T.}~\bibnamefont {Liu}},\ and\ \bibinfo {author} {\bibfnamefont {F.}~\bibnamefont {Nori}},\ }\bibfield  {title} {\bibinfo {title} {Topological quantum phase transitions retrieved through unsupervised machine learning},\ }\href {https://doi.org/10.1103/PhysRevB.102.134213} {\bibfield  {journal} {\bibinfo  {journal} {Physical Review B}\ }\textbf {\bibinfo {volume} {102}},\ \bibinfo {pages} {134213} (\bibinfo {year} {2020})}\BibitemShut {NoStop}%
\bibitem [{\citenamefont {Yu}\ \emph {et~al.}(2024)\citenamefont {Yu}, \citenamefont {Zhang}, \citenamefont {Shen},\ and\ \citenamefont {Deng}}]{yuUnsupervisedLearningInteracting2024}%
  \BibitemOpen
  \bibfield  {author} {\bibinfo {author} {\bibfnamefont {L.-W.}\ \bibnamefont {Yu}}, \bibinfo {author} {\bibfnamefont {S.-Y.}\ \bibnamefont {Zhang}}, \bibinfo {author} {\bibfnamefont {P.-X.}\ \bibnamefont {Shen}},\ and\ \bibinfo {author} {\bibfnamefont {D.-L.}\ \bibnamefont {Deng}},\ }\bibfield  {title} {\bibinfo {title} {Unsupervised learning of interacting topological phases from experimental observables},\ }\href {https://doi.org/10.1016/j.fmre.2022.12.016} {\bibfield  {journal} {\bibinfo  {journal} {Fundamental Research}\ }\textbf {\bibinfo {volume} {4}},\ \bibinfo {pages} {1086} (\bibinfo {year} {2024})}\BibitemShut {NoStop}%
\bibitem [{\citenamefont {Ziv}\ \emph {et~al.}(2025)\citenamefont {Ziv}, \citenamefont {Wei}, \citenamefont {{Rubio-Abadal}}, \citenamefont {Adler}, \citenamefont {Keselman}, \citenamefont {Lustig}, \citenamefont {Talmon}, \citenamefont {Zeiher}, \citenamefont {Bloch},\ and\ \citenamefont {Segev}}]{zivUnsupervisedMachineLearning2025}%
  \BibitemOpen
  \bibfield  {author} {\bibinfo {author} {\bibfnamefont {R.}~\bibnamefont {Ziv}}, \bibinfo {author} {\bibfnamefont {D.}~\bibnamefont {Wei}}, \bibinfo {author} {\bibfnamefont {A.}~\bibnamefont {{Rubio-Abadal}}}, \bibinfo {author} {\bibfnamefont {D.}~\bibnamefont {Adler}}, \bibinfo {author} {\bibfnamefont {A.}~\bibnamefont {Keselman}}, \bibinfo {author} {\bibfnamefont {E.}~\bibnamefont {Lustig}}, \bibinfo {author} {\bibfnamefont {R.}~\bibnamefont {Talmon}}, \bibinfo {author} {\bibfnamefont {J.}~\bibnamefont {Zeiher}}, \bibinfo {author} {\bibfnamefont {I.}~\bibnamefont {Bloch}},\ and\ \bibinfo {author} {\bibfnamefont {M.}~\bibnamefont {Segev}},\ }\href {https://doi.org/10.48550/arXiv.2512.01091} {\bibinfo {title} {Unsupervised {{Machine Learning}} for {{Experimental Detection}} of {{Quantum-Many-Body Phase Transitions}}}} (\bibinfo {year} {2025}),\ \Eprint {https://arxiv.org/abs/2512.01091} {arXiv:2512.01091 [quant-ph]} \BibitemShut {NoStop}%
\bibitem [{\citenamefont {Kuo}\ and\ \citenamefont {Dehghani}(2022)}]{kuoUnsupervisedLearningInteracting2022}%
  \BibitemOpen
  \bibfield  {author} {\bibinfo {author} {\bibfnamefont {E.-J.}\ \bibnamefont {Kuo}}\ and\ \bibinfo {author} {\bibfnamefont {H.}~\bibnamefont {Dehghani}},\ }\bibfield  {title} {\bibinfo {title} {Unsupervised learning of interacting topological and symmetry-breaking phase transitions},\ }\href {https://doi.org/10.1103/PhysRevB.105.235136} {\bibfield  {journal} {\bibinfo  {journal} {Physical Review B}\ }\textbf {\bibinfo {volume} {105}},\ \bibinfo {pages} {235136} (\bibinfo {year} {2022})}\BibitemShut {NoStop}%
\bibitem [{\citenamefont {Baier}\ \emph {et~al.}(2016)\citenamefont {Baier}, \citenamefont {Mark}, \citenamefont {Petter}, \citenamefont {Aikawa}, \citenamefont {Chomaz}, \citenamefont {Cai}, \citenamefont {Baranov}, \citenamefont {Zoller},\ and\ \citenamefont {Ferlaino}}]{ebhmFrancescaSci2016}%
  \BibitemOpen
  \bibfield  {author} {\bibinfo {author} {\bibfnamefont {S.}~\bibnamefont {Baier}}, \bibinfo {author} {\bibfnamefont {M.~J.}\ \bibnamefont {Mark}}, \bibinfo {author} {\bibfnamefont {D.}~\bibnamefont {Petter}}, \bibinfo {author} {\bibfnamefont {K.}~\bibnamefont {Aikawa}}, \bibinfo {author} {\bibfnamefont {L.}~\bibnamefont {Chomaz}}, \bibinfo {author} {\bibfnamefont {Z.}~\bibnamefont {Cai}}, \bibinfo {author} {\bibfnamefont {M.}~\bibnamefont {Baranov}}, \bibinfo {author} {\bibfnamefont {P.}~\bibnamefont {Zoller}},\ and\ \bibinfo {author} {\bibfnamefont {F.}~\bibnamefont {Ferlaino}},\ }\bibfield  {title} {\bibinfo {title} {Extended {{Bose-Hubbard}} models with ultracold magnetic atoms},\ }\href {https://doi.org/10.1126/science.aac9812} {\bibfield  {journal} {\bibinfo  {journal} {Science}\ }\textbf {\bibinfo {volume} {352}},\ \bibinfo {pages} {201} (\bibinfo {year} {2016})}\BibitemShut {NoStop}%
\bibitem [{\citenamefont {Fersterer}\ \emph {et~al.}(2019)\citenamefont {Fersterer}, \citenamefont {{Safavi-Naini}}, \citenamefont {Zhu}, \citenamefont {Gabardos}, \citenamefont {Lepoutre}, \citenamefont {Vernac}, \citenamefont {{Laburthe-Tolra}}, \citenamefont {Blakie},\ and\ \citenamefont {Rey}}]{ferstererDynamicsItinerantSpin32019}%
  \BibitemOpen
  \bibfield  {author} {\bibinfo {author} {\bibfnamefont {P.}~\bibnamefont {Fersterer}}, \bibinfo {author} {\bibfnamefont {A.}~\bibnamefont {{Safavi-Naini}}}, \bibinfo {author} {\bibfnamefont {B.}~\bibnamefont {Zhu}}, \bibinfo {author} {\bibfnamefont {L.}~\bibnamefont {Gabardos}}, \bibinfo {author} {\bibfnamefont {S.}~\bibnamefont {Lepoutre}}, \bibinfo {author} {\bibfnamefont {L.}~\bibnamefont {Vernac}}, \bibinfo {author} {\bibfnamefont {B.}~\bibnamefont {{Laburthe-Tolra}}}, \bibinfo {author} {\bibfnamefont {P.~B.}\ \bibnamefont {Blakie}},\ and\ \bibinfo {author} {\bibfnamefont {A.~M.}\ \bibnamefont {Rey}},\ }\bibfield  {title} {\bibinfo {title} {Dynamics of an itinerant spin-3 atomic dipolar gas in an optical lattice},\ }\href {https://doi.org/10.1103/PhysRevA.100.033609} {\bibfield  {journal} {\bibinfo  {journal} {Physical Review A}\ }\textbf {\bibinfo {volume} {100}},\ \bibinfo {pages} {033609} (\bibinfo {year} {2019})}\BibitemShut {NoStop}%
\bibitem [{\citenamefont {Su}\ \emph {et~al.}(2023)\citenamefont {Su}, \citenamefont {Douglas}, \citenamefont {Szurek}, \citenamefont {Groth}, \citenamefont {Ozturk}, \citenamefont {Krahn}, \citenamefont {H{\'e}bert}, \citenamefont {Phelps}, \citenamefont {Ebadi}, \citenamefont {Dickerson}, \citenamefont {Ferlaino}, \citenamefont {Markovi{\'c}},\ and\ \citenamefont {Greiner}}]{suDipolarQuantumSolids2023}%
  \BibitemOpen
  \bibfield  {author} {\bibinfo {author} {\bibfnamefont {L.}~\bibnamefont {Su}}, \bibinfo {author} {\bibfnamefont {A.}~\bibnamefont {Douglas}}, \bibinfo {author} {\bibfnamefont {M.}~\bibnamefont {Szurek}}, \bibinfo {author} {\bibfnamefont {R.}~\bibnamefont {Groth}}, \bibinfo {author} {\bibfnamefont {S.~F.}\ \bibnamefont {Ozturk}}, \bibinfo {author} {\bibfnamefont {A.}~\bibnamefont {Krahn}}, \bibinfo {author} {\bibfnamefont {A.~H.}\ \bibnamefont {H{\'e}bert}}, \bibinfo {author} {\bibfnamefont {G.~A.}\ \bibnamefont {Phelps}}, \bibinfo {author} {\bibfnamefont {S.}~\bibnamefont {Ebadi}}, \bibinfo {author} {\bibfnamefont {S.}~\bibnamefont {Dickerson}}, \bibinfo {author} {\bibfnamefont {F.}~\bibnamefont {Ferlaino}}, \bibinfo {author} {\bibfnamefont {O.}~\bibnamefont {Markovi{\'c}}},\ and\ \bibinfo {author} {\bibfnamefont {M.}~\bibnamefont {Greiner}},\ }\bibfield  {title} {\bibinfo {title} {Dipolar quantum solids emerging in a {{Hubbard}} quantum simulator},\ }\href {https://doi.org/10.1038/s41586-023-06614-3}
  {\bibfield  {journal} {\bibinfo  {journal} {Nature}\ }\textbf {\bibinfo {volume} {622}},\ \bibinfo {pages} {724} (\bibinfo {year} {2023})}\BibitemShut {NoStop}%
\bibitem [{\citenamefont {Dalla~Torre}\ \emph {et~al.}(2006)\citenamefont {Dalla~Torre}, \citenamefont {Berg},\ and\ \citenamefont {Altman}}]{dallatorreHiddenOrder1D2006}%
  \BibitemOpen
  \bibfield  {author} {\bibinfo {author} {\bibfnamefont {E.~G.}\ \bibnamefont {Dalla~Torre}}, \bibinfo {author} {\bibfnamefont {E.}~\bibnamefont {Berg}},\ and\ \bibinfo {author} {\bibfnamefont {E.}~\bibnamefont {Altman}},\ }\bibfield  {title} {\bibinfo {title} {Hidden {{Order}} in {{1D Bose Insulators}}},\ }\href {https://doi.org/10.1103/PhysRevLett.97.260401} {\bibfield  {journal} {\bibinfo  {journal} {Physical Review Letters}\ }\textbf {\bibinfo {volume} {97}},\ \bibinfo {pages} {260401} (\bibinfo {year} {2006})}\BibitemShut {NoStop}%
\bibitem [{\citenamefont {Ejima}\ \emph {et~al.}(2014)\citenamefont {Ejima}, \citenamefont {Lange},\ and\ \citenamefont {Fehske}}]{ejimaSpectralEntanglementProperties2014}%
  \BibitemOpen
  \bibfield  {author} {\bibinfo {author} {\bibfnamefont {S.}~\bibnamefont {Ejima}}, \bibinfo {author} {\bibfnamefont {F.}~\bibnamefont {Lange}},\ and\ \bibinfo {author} {\bibfnamefont {H.}~\bibnamefont {Fehske}},\ }\bibfield  {title} {\bibinfo {title} {Spectral and {{Entanglement Properties}} of the {{Bosonic Haldane Insulator}}},\ }\href {https://doi.org/10.1103/PhysRevLett.113.020401} {\bibfield  {journal} {\bibinfo  {journal} {Physical Review Letters}\ }\textbf {\bibinfo {volume} {113}},\ \bibinfo {pages} {020401} (\bibinfo {year} {2014})}\BibitemShut {NoStop}%
\bibitem [{\citenamefont {Berg}\ \emph {et~al.}(2008)\citenamefont {Berg}, \citenamefont {Dalla~Torre}, \citenamefont {Giamarchi},\ and\ \citenamefont {Altman}}]{bergRiseFallHidden2008}%
  \BibitemOpen
  \bibfield  {author} {\bibinfo {author} {\bibfnamefont {E.}~\bibnamefont {Berg}}, \bibinfo {author} {\bibfnamefont {E.~G.}\ \bibnamefont {Dalla~Torre}}, \bibinfo {author} {\bibfnamefont {T.}~\bibnamefont {Giamarchi}},\ and\ \bibinfo {author} {\bibfnamefont {E.}~\bibnamefont {Altman}},\ }\bibfield  {title} {\bibinfo {title} {Rise and fall of hidden string order of lattice bosons},\ }\href {https://doi.org/10.1103/PhysRevB.77.245119} {\bibfield  {journal} {\bibinfo  {journal} {Physical Review B}\ }\textbf {\bibinfo {volume} {77}},\ \bibinfo {pages} {245119} (\bibinfo {year} {2008})}\BibitemShut {NoStop}%
\bibitem [{\citenamefont {Rossini}\ and\ \citenamefont {Fazio}(2012)}]{rossiniPhaseDiagramExtended2012}%
  \BibitemOpen
  \bibfield  {author} {\bibinfo {author} {\bibfnamefont {D.}~\bibnamefont {Rossini}}\ and\ \bibinfo {author} {\bibfnamefont {R.}~\bibnamefont {Fazio}},\ }\bibfield  {title} {\bibinfo {title} {Phase diagram of the extended {{Bose}}--{{Hubbard}} model},\ }\href {https://doi.org/10.1088/1367-2630/14/6/065012} {\bibfield  {journal} {\bibinfo  {journal} {New Journal of Physics}\ }\textbf {\bibinfo {volume} {14}},\ \bibinfo {pages} {065012} (\bibinfo {year} {2012})}\BibitemShut {NoStop}%
\bibitem [{\citenamefont {Fu}(2011)}]{fuTopologicalCrystallineInsulators2011}%
  \BibitemOpen
  \bibfield  {author} {\bibinfo {author} {\bibfnamefont {L.}~\bibnamefont {Fu}},\ }\bibfield  {title} {\bibinfo {title} {Topological {{Crystalline Insulators}}},\ }\href {https://doi.org/10.1103/PhysRevLett.106.106802} {\bibfield  {journal} {\bibinfo  {journal} {Physical Review Letters}\ }\textbf {\bibinfo {volume} {106}},\ \bibinfo {pages} {106802} (\bibinfo {year} {2011})}\BibitemShut {NoStop}%
\bibitem [{\citenamefont {Hughes}\ \emph {et~al.}(2011)\citenamefont {Hughes}, \citenamefont {Prodan},\ and\ \citenamefont {Bernevig}}]{hughesInversionsymmetricTopologicalInsulators2011}%
  \BibitemOpen
  \bibfield  {author} {\bibinfo {author} {\bibfnamefont {T.~L.}\ \bibnamefont {Hughes}}, \bibinfo {author} {\bibfnamefont {E.}~\bibnamefont {Prodan}},\ and\ \bibinfo {author} {\bibfnamefont {B.~A.}\ \bibnamefont {Bernevig}},\ }\bibfield  {title} {\bibinfo {title} {Inversion-symmetric topological insulators},\ }\href {https://doi.org/10.1103/PhysRevB.83.245132} {\bibfield  {journal} {\bibinfo  {journal} {Physical Review B}\ }\textbf {\bibinfo {volume} {83}},\ \bibinfo {pages} {245132} (\bibinfo {year} {2011})}\BibitemShut {NoStop}%
\bibitem [{\citenamefont {Turner}\ \emph {et~al.}(2010)\citenamefont {Turner}, \citenamefont {Zhang},\ and\ \citenamefont {Vishwanath}}]{turnerEntanglementInversionSymmetry2010}%
  \BibitemOpen
  \bibfield  {author} {\bibinfo {author} {\bibfnamefont {A.~M.}\ \bibnamefont {Turner}}, \bibinfo {author} {\bibfnamefont {Y.}~\bibnamefont {Zhang}},\ and\ \bibinfo {author} {\bibfnamefont {A.}~\bibnamefont {Vishwanath}},\ }\bibfield  {title} {\bibinfo {title} {Entanglement and inversion symmetry in topological insulators},\ }\href {https://doi.org/10.1103/PhysRevB.82.241102} {\bibfield  {journal} {\bibinfo  {journal} {Physical Review B}\ }\textbf {\bibinfo {volume} {82}},\ \bibinfo {pages} {241102} (\bibinfo {year} {2010})}\BibitemShut {NoStop}%
\bibitem [{\citenamefont {Rispoli}\ \emph {et~al.}(2019)\citenamefont {Rispoli}, \citenamefont {Lukin}, \citenamefont {Schittko}, \citenamefont {Kim}, \citenamefont {Tai}, \citenamefont {L{\'e}onard},\ and\ \citenamefont {Greiner}}]{rispoliQuantumCriticalBehaviour2019}%
  \BibitemOpen
  \bibfield  {author} {\bibinfo {author} {\bibfnamefont {M.}~\bibnamefont {Rispoli}}, \bibinfo {author} {\bibfnamefont {A.}~\bibnamefont {Lukin}}, \bibinfo {author} {\bibfnamefont {R.}~\bibnamefont {Schittko}}, \bibinfo {author} {\bibfnamefont {S.}~\bibnamefont {Kim}}, \bibinfo {author} {\bibfnamefont {M.~E.}\ \bibnamefont {Tai}}, \bibinfo {author} {\bibfnamefont {J.}~\bibnamefont {L{\'e}onard}},\ and\ \bibinfo {author} {\bibfnamefont {M.}~\bibnamefont {Greiner}},\ }\bibfield  {title} {\bibinfo {title} {Quantum critical behaviour at the many-body localization transition},\ }\href {https://doi.org/10.1038/s41586-019-1527-2} {\bibfield  {journal} {\bibinfo  {journal} {Nature}\ }\textbf {\bibinfo {volume} {573}},\ \bibinfo {pages} {385} (\bibinfo {year} {2019})}\BibitemShut {NoStop}%
\bibitem [{\citenamefont {Basko}\ \emph {et~al.}(2006)\citenamefont {Basko}, \citenamefont {Aleiner},\ and\ \citenamefont {Altshuler}}]{baskoMetalInsulatorTransition2006}%
  \BibitemOpen
  \bibfield  {author} {\bibinfo {author} {\bibfnamefont {D.~M.}\ \bibnamefont {Basko}}, \bibinfo {author} {\bibfnamefont {I.~L.}\ \bibnamefont {Aleiner}},\ and\ \bibinfo {author} {\bibfnamefont {B.~L.}\ \bibnamefont {Altshuler}},\ }\bibfield  {title} {\bibinfo {title} {Metal--insulator transition in a weakly interacting many-electron system with localized single-particle states},\ }\href {https://doi.org/10.1016/j.aop.2005.11.014} {\bibfield  {journal} {\bibinfo  {journal} {Annals of Physics}\ }\textbf {\bibinfo {volume} {321}},\ \bibinfo {pages} {1126} (\bibinfo {year} {2006})}\BibitemShut {NoStop}%
\bibitem [{\citenamefont {Abanin}\ and\ \citenamefont {Papi{\'c}}(2017)}]{abaninRecentProgressManybody2017}%
  \BibitemOpen
  \bibfield  {author} {\bibinfo {author} {\bibfnamefont {D.~A.}\ \bibnamefont {Abanin}}\ and\ \bibinfo {author} {\bibfnamefont {Z.}~\bibnamefont {Papi{\'c}}},\ }\bibfield  {title} {\bibinfo {title} {Recent progress in many-body localization},\ }\href {https://doi.org/10.1002/andp.201700169} {\bibfield  {journal} {\bibinfo  {journal} {Annalen der Physik}\ }\textbf {\bibinfo {volume} {529}},\ \bibinfo {pages} {1700169} (\bibinfo {year} {2017})}\BibitemShut {NoStop}%
\bibitem [{\citenamefont {Nandkishore}\ and\ \citenamefont {Huse}(2015)}]{nandkishoreManyBodyLocalizationThermalization2015}%
  \BibitemOpen
  \bibfield  {author} {\bibinfo {author} {\bibfnamefont {R.}~\bibnamefont {Nandkishore}}\ and\ \bibinfo {author} {\bibfnamefont {D.~A.}\ \bibnamefont {Huse}},\ }\bibfield  {title} {\bibinfo {title} {Many-{{Body Localization}} and {{Thermalization}} in {{Quantum Statistical Mechanics}}},\ }\href {https://doi.org/10.1146/annurev-conmatphys-031214-014726} {\bibfield  {journal} {\bibinfo  {journal} {Annual Review of Condensed Matter Physics}\ }\textbf {\bibinfo {volume} {6}},\ \bibinfo {pages} {15} (\bibinfo {year} {2015})}\BibitemShut {NoStop}%
\bibitem [{\citenamefont {Huse}\ \emph {et~al.}(2014)\citenamefont {Huse}, \citenamefont {Nandkishore},\ and\ \citenamefont {Oganesyan}}]{husePhenomenologyFullyManybodylocalized2014}%
  \BibitemOpen
  \bibfield  {author} {\bibinfo {author} {\bibfnamefont {D.~A.}\ \bibnamefont {Huse}}, \bibinfo {author} {\bibfnamefont {R.}~\bibnamefont {Nandkishore}},\ and\ \bibinfo {author} {\bibfnamefont {V.}~\bibnamefont {Oganesyan}},\ }\bibfield  {title} {\bibinfo {title} {Phenomenology of fully many-body-localized systems},\ }\href {https://doi.org/10.1103/PhysRevB.90.174202} {\bibfield  {journal} {\bibinfo  {journal} {Physical Review B}\ }\textbf {\bibinfo {volume} {90}},\ \bibinfo {pages} {174202} (\bibinfo {year} {2014})}\BibitemShut {NoStop}%
\bibitem [{\citenamefont {Wahl}\ \emph {et~al.}(2019)\citenamefont {Wahl}, \citenamefont {Pal},\ and\ \citenamefont {Simon}}]{wahlSignaturesManybodyLocalized2019}%
  \BibitemOpen
  \bibfield  {author} {\bibinfo {author} {\bibfnamefont {T.~B.}\ \bibnamefont {Wahl}}, \bibinfo {author} {\bibfnamefont {A.}~\bibnamefont {Pal}},\ and\ \bibinfo {author} {\bibfnamefont {S.~H.}\ \bibnamefont {Simon}},\ }\bibfield  {title} {\bibinfo {title} {Signatures of the many-body localized regime in two dimensions},\ }\href {https://doi.org/10.1038/s41567-018-0339-x} {\bibfield  {journal} {\bibinfo  {journal} {Nature Physics}\ }\textbf {\bibinfo {volume} {15}},\ \bibinfo {pages} {164} (\bibinfo {year} {2019})}\BibitemShut {NoStop}%
\bibitem [{\citenamefont {Alet}\ and\ \citenamefont {Laflorencie}(2018)}]{aletManybodyLocalizationIntroduction2018}%
  \BibitemOpen
  \bibfield  {author} {\bibinfo {author} {\bibfnamefont {F.}~\bibnamefont {Alet}}\ and\ \bibinfo {author} {\bibfnamefont {N.}~\bibnamefont {Laflorencie}},\ }\bibfield  {title} {\bibinfo {title} {Many-body localization: {{An}} introduction and selected topics},\ }\href {https://doi.org/10.1016/j.crhy.2018.03.003} {\bibfield  {journal} {\bibinfo  {journal} {Comptes Rendus. Physique}\ }\textbf {\bibinfo {volume} {19}},\ \bibinfo {pages} {498} (\bibinfo {year} {2018})}\BibitemShut {NoStop}%
\bibitem [{\citenamefont {Sierant}\ and\ \citenamefont {Zakrzewski}(2018)}]{sierantManybodyLocalizationBosons2018}%
  \BibitemOpen
  \bibfield  {author} {\bibinfo {author} {\bibfnamefont {P.}~\bibnamefont {Sierant}}\ and\ \bibinfo {author} {\bibfnamefont {J.}~\bibnamefont {Zakrzewski}},\ }\bibfield  {title} {\bibinfo {title} {Many-body localization of bosons in optical lattices},\ }\href {https://doi.org/10.1088/1367-2630/aabb17} {\bibfield  {journal} {\bibinfo  {journal} {New Journal of Physics}\ }\textbf {\bibinfo {volume} {20}},\ \bibinfo {pages} {043032} (\bibinfo {year} {2018})}\BibitemShut {NoStop}%
\bibitem [{\citenamefont {Yao}\ and\ \citenamefont {Zakrzewski}(2020)}]{yaoManybodyLocalizationBoseHubbard2020}%
  \BibitemOpen
  \bibfield  {author} {\bibinfo {author} {\bibfnamefont {R.}~\bibnamefont {Yao}}\ and\ \bibinfo {author} {\bibfnamefont {J.}~\bibnamefont {Zakrzewski}},\ }\bibfield  {title} {\bibinfo {title} {Many-body localization in the {{Bose-Hubbard}} model: {{Evidence}} for mobility edge},\ }\href {https://doi.org/10.1103/PhysRevB.102.014310} {\bibfield  {journal} {\bibinfo  {journal} {Physical Review B}\ }\textbf {\bibinfo {volume} {102}},\ \bibinfo {pages} {014310} (\bibinfo {year} {2020})}\BibitemShut {NoStop}%
\bibitem [{\citenamefont {L{\"u}schen}\ \emph {et~al.}(2017)\citenamefont {L{\"u}schen}, \citenamefont {Bordia}, \citenamefont {Scherg}, \citenamefont {Alet}, \citenamefont {Altman}, \citenamefont {Schneider},\ and\ \citenamefont {Bloch}}]{luschenObservationSlowDynamics2017}%
  \BibitemOpen
  \bibfield  {author} {\bibinfo {author} {\bibfnamefont {H.~P.}\ \bibnamefont {L{\"u}schen}}, \bibinfo {author} {\bibfnamefont {P.}~\bibnamefont {Bordia}}, \bibinfo {author} {\bibfnamefont {S.}~\bibnamefont {Scherg}}, \bibinfo {author} {\bibfnamefont {F.}~\bibnamefont {Alet}}, \bibinfo {author} {\bibfnamefont {E.}~\bibnamefont {Altman}}, \bibinfo {author} {\bibfnamefont {U.}~\bibnamefont {Schneider}},\ and\ \bibinfo {author} {\bibfnamefont {I.}~\bibnamefont {Bloch}},\ }\bibfield  {title} {\bibinfo {title} {Observation of {{Slow Dynamics}} near the {{Many-Body Localization Transition}} in {{One-Dimensional Quasiperiodic Systems}}},\ }\href {https://doi.org/10.1103/PhysRevLett.119.260401} {\bibfield  {journal} {\bibinfo  {journal} {Physical Review Letters}\ }\textbf {\bibinfo {volume} {119}},\ \bibinfo {pages} {260401} (\bibinfo {year} {2017})}\BibitemShut {NoStop}%
\bibitem [{\citenamefont {Schreiber}\ \emph {et~al.}(2015)\citenamefont {Schreiber}, \citenamefont {Hodgman}, \citenamefont {Bordia}, \citenamefont {L{\"u}schen}, \citenamefont {Fischer}, \citenamefont {Vosk}, \citenamefont {Altman}, \citenamefont {Schneider},\ and\ \citenamefont {Bloch}}]{schreiberObservationManybodyLocalization2015}%
  \BibitemOpen
  \bibfield  {author} {\bibinfo {author} {\bibfnamefont {M.}~\bibnamefont {Schreiber}}, \bibinfo {author} {\bibfnamefont {S.~S.}\ \bibnamefont {Hodgman}}, \bibinfo {author} {\bibfnamefont {P.}~\bibnamefont {Bordia}}, \bibinfo {author} {\bibfnamefont {H.~P.}\ \bibnamefont {L{\"u}schen}}, \bibinfo {author} {\bibfnamefont {M.~H.}\ \bibnamefont {Fischer}}, \bibinfo {author} {\bibfnamefont {R.}~\bibnamefont {Vosk}}, \bibinfo {author} {\bibfnamefont {E.}~\bibnamefont {Altman}}, \bibinfo {author} {\bibfnamefont {U.}~\bibnamefont {Schneider}},\ and\ \bibinfo {author} {\bibfnamefont {I.}~\bibnamefont {Bloch}},\ }\bibfield  {title} {\bibinfo {title} {Observation of many-body localization of interacting fermions in a quasirandom optical lattice},\ }\href {https://doi.org/10.1126/science.aaa7432} {\bibfield  {journal} {\bibinfo  {journal} {Science}\ }\textbf {\bibinfo {volume} {349}},\ \bibinfo {pages} {842} (\bibinfo {year} {2015})}\BibitemShut {NoStop}%
\bibitem [{\citenamefont {Choi}\ \emph {et~al.}(2016)\citenamefont {Choi}, \citenamefont {Hild}, \citenamefont {Zeiher}, \citenamefont {Schau{\ss}}, \citenamefont {{Rubio-Abadal}}, \citenamefont {Yefsah}, \citenamefont {Khemani}, \citenamefont {Huse}, \citenamefont {Bloch},\ and\ \citenamefont {Gross}}]{choiExploringManybodyLocalization2016}%
  \BibitemOpen
  \bibfield  {author} {\bibinfo {author} {\bibfnamefont {J.-y.}\ \bibnamefont {Choi}}, \bibinfo {author} {\bibfnamefont {S.}~\bibnamefont {Hild}}, \bibinfo {author} {\bibfnamefont {J.}~\bibnamefont {Zeiher}}, \bibinfo {author} {\bibfnamefont {P.}~\bibnamefont {Schau{\ss}}}, \bibinfo {author} {\bibfnamefont {A.}~\bibnamefont {{Rubio-Abadal}}}, \bibinfo {author} {\bibfnamefont {T.}~\bibnamefont {Yefsah}}, \bibinfo {author} {\bibfnamefont {V.}~\bibnamefont {Khemani}}, \bibinfo {author} {\bibfnamefont {D.~A.}\ \bibnamefont {Huse}}, \bibinfo {author} {\bibfnamefont {I.}~\bibnamefont {Bloch}},\ and\ \bibinfo {author} {\bibfnamefont {C.}~\bibnamefont {Gross}},\ }\bibfield  {title} {\bibinfo {title} {Exploring the many-body localization transition in two dimensions},\ }\href {https://doi.org/10.1126/science.aaf8834} {\bibfield  {journal} {\bibinfo  {journal} {Science}\ }\textbf {\bibinfo {volume} {352}},\ \bibinfo {pages} {1547} (\bibinfo {year} {2016})},\ \Eprint {https://arxiv.org/abs/1604.04178} {arXiv:1604.04178
  [cond-mat]} \BibitemShut {NoStop}%
\bibitem [{\citenamefont {Kondov}\ \emph {et~al.}(2015)\citenamefont {Kondov}, \citenamefont {McGehee}, \citenamefont {Xu},\ and\ \citenamefont {DeMarco}}]{kondovDisorderInducedLocalizationStrongly2015}%
  \BibitemOpen
  \bibfield  {author} {\bibinfo {author} {\bibfnamefont {S.~S.}\ \bibnamefont {Kondov}}, \bibinfo {author} {\bibfnamefont {W.~R.}\ \bibnamefont {McGehee}}, \bibinfo {author} {\bibfnamefont {W.}~\bibnamefont {Xu}},\ and\ \bibinfo {author} {\bibfnamefont {B.}~\bibnamefont {DeMarco}},\ }\bibfield  {title} {\bibinfo {title} {Disorder-{{Induced Localization}} in a {{Strongly Correlated Atomic Hubbard Gas}}},\ }\href {https://doi.org/10.1103/PhysRevLett.114.083002} {\bibfield  {journal} {\bibinfo  {journal} {Physical Review Letters}\ }\textbf {\bibinfo {volume} {114}},\ \bibinfo {pages} {083002} (\bibinfo {year} {2015})}\BibitemShut {NoStop}%
\bibitem [{\citenamefont {Pal}\ and\ \citenamefont {Huse}(2010)}]{palManybodyLocalizationPhase2010}%
  \BibitemOpen
  \bibfield  {author} {\bibinfo {author} {\bibfnamefont {A.}~\bibnamefont {Pal}}\ and\ \bibinfo {author} {\bibfnamefont {D.~A.}\ \bibnamefont {Huse}},\ }\bibfield  {title} {\bibinfo {title} {Many-body localization phase transition},\ }\href {https://doi.org/10.1103/PhysRevB.82.174411} {\bibfield  {journal} {\bibinfo  {journal} {Physical Review B}\ }\textbf {\bibinfo {volume} {82}},\ \bibinfo {pages} {174411} (\bibinfo {year} {2010})}\BibitemShut {NoStop}%
\bibitem [{\citenamefont {Khemani}\ \emph {et~al.}(2017)\citenamefont {Khemani}, \citenamefont {Lim}, \citenamefont {Sheng},\ and\ \citenamefont {Huse}}]{khemaniCriticalPropertiesManyBody2017}%
  \BibitemOpen
  \bibfield  {author} {\bibinfo {author} {\bibfnamefont {V.}~\bibnamefont {Khemani}}, \bibinfo {author} {\bibfnamefont {S.~P.}\ \bibnamefont {Lim}}, \bibinfo {author} {\bibfnamefont {D.~N.}\ \bibnamefont {Sheng}},\ and\ \bibinfo {author} {\bibfnamefont {D.~A.}\ \bibnamefont {Huse}},\ }\bibfield  {title} {\bibinfo {title} {Critical {{Properties}} of the {{Many-Body Localization Transition}}},\ }\href {https://doi.org/10.1103/PhysRevX.7.021013} {\bibfield  {journal} {\bibinfo  {journal} {Physical Review X}\ }\textbf {\bibinfo {volume} {7}},\ \bibinfo {pages} {021013} (\bibinfo {year} {2017})}\BibitemShut {NoStop}%
\bibitem [{\citenamefont {Luitz}\ \emph {et~al.}(2015)\citenamefont {Luitz}, \citenamefont {Laflorencie},\ and\ \citenamefont {Alet}}]{luitzManybodyLocalizationEdge2015}%
  \BibitemOpen
  \bibfield  {author} {\bibinfo {author} {\bibfnamefont {D.~J.}\ \bibnamefont {Luitz}}, \bibinfo {author} {\bibfnamefont {N.}~\bibnamefont {Laflorencie}},\ and\ \bibinfo {author} {\bibfnamefont {F.}~\bibnamefont {Alet}},\ }\bibfield  {title} {\bibinfo {title} {Many-body localization edge in the random-field {{Heisenberg}} chain},\ }\href {https://doi.org/10.1103/PhysRevB.91.081103} {\bibfield  {journal} {\bibinfo  {journal} {Physical Review B}\ }\textbf {\bibinfo {volume} {91}},\ \bibinfo {pages} {081103} (\bibinfo {year} {2015})}\BibitemShut {NoStop}%
\bibitem [{\citenamefont {Guo}\ \emph {et~al.}(2021)\citenamefont {Guo}, \citenamefont {Cheng}, \citenamefont {Sun}, \citenamefont {Song}, \citenamefont {Li}, \citenamefont {Wang}, \citenamefont {Ren}, \citenamefont {Dong}, \citenamefont {Zheng}, \citenamefont {Zhang}, \citenamefont {Mondaini}, \citenamefont {Fan},\ and\ \citenamefont {Wang}}]{guoObservationEnergyresolvedManybody2021}%
  \BibitemOpen
  \bibfield  {author} {\bibinfo {author} {\bibfnamefont {Q.}~\bibnamefont {Guo}}, \bibinfo {author} {\bibfnamefont {C.}~\bibnamefont {Cheng}}, \bibinfo {author} {\bibfnamefont {Z.-H.}\ \bibnamefont {Sun}}, \bibinfo {author} {\bibfnamefont {Z.}~\bibnamefont {Song}}, \bibinfo {author} {\bibfnamefont {H.}~\bibnamefont {Li}}, \bibinfo {author} {\bibfnamefont {Z.}~\bibnamefont {Wang}}, \bibinfo {author} {\bibfnamefont {W.}~\bibnamefont {Ren}}, \bibinfo {author} {\bibfnamefont {H.}~\bibnamefont {Dong}}, \bibinfo {author} {\bibfnamefont {D.}~\bibnamefont {Zheng}}, \bibinfo {author} {\bibfnamefont {Y.-R.}\ \bibnamefont {Zhang}}, \bibinfo {author} {\bibfnamefont {R.}~\bibnamefont {Mondaini}}, \bibinfo {author} {\bibfnamefont {H.}~\bibnamefont {Fan}},\ and\ \bibinfo {author} {\bibfnamefont {H.}~\bibnamefont {Wang}},\ }\bibfield  {title} {\bibinfo {title} {Observation of energy-resolved many-body localization},\ }\href {https://doi.org/10.1038/s41567-020-1035-1} {\bibfield  {journal} {\bibinfo  {journal} {Nature
  Physics}\ }\textbf {\bibinfo {volume} {17}},\ \bibinfo {pages} {234} (\bibinfo {year} {2021})}\BibitemShut {NoStop}%
\bibitem [{\citenamefont {Orell}\ \emph {et~al.}(2019)\citenamefont {Orell}, \citenamefont {Michailidis}, \citenamefont {Serbyn},\ and\ \citenamefont {Silveri}}]{orellProbingManybodyLocalization2019}%
  \BibitemOpen
  \bibfield  {author} {\bibinfo {author} {\bibfnamefont {T.}~\bibnamefont {Orell}}, \bibinfo {author} {\bibfnamefont {A.~A.}\ \bibnamefont {Michailidis}}, \bibinfo {author} {\bibfnamefont {M.}~\bibnamefont {Serbyn}},\ and\ \bibinfo {author} {\bibfnamefont {M.}~\bibnamefont {Silveri}},\ }\bibfield  {title} {\bibinfo {title} {Probing the many-body localization phase transition with superconducting circuits},\ }\href {https://doi.org/10.1103/PhysRevB.100.134504} {\bibfield  {journal} {\bibinfo  {journal} {Physical Review B}\ }\textbf {\bibinfo {volume} {100}},\ \bibinfo {pages} {134504} (\bibinfo {year} {2019})}\BibitemShut {NoStop}%
\bibitem [{\citenamefont {Coifman}\ and\ \citenamefont {Lafon}(2006)}]{coifmanDiffusionMaps2006}%
  \BibitemOpen
  \bibfield  {author} {\bibinfo {author} {\bibfnamefont {R.~R.}\ \bibnamefont {Coifman}}\ and\ \bibinfo {author} {\bibfnamefont {S.}~\bibnamefont {Lafon}},\ }\bibfield  {title} {\bibinfo {title} {Diffusion maps},\ }\href {https://doi.org/10.1016/j.acha.2006.04.006} {\bibfield  {journal} {\bibinfo  {journal} {Applied and Computational Harmonic Analysis}\ }\textbf {\bibinfo {volume} {21}},\ \bibinfo {pages} {5} (\bibinfo {year} {2006})}\BibitemShut {NoStop}%
\bibitem [{\citenamefont {Nadler}\ \emph {et~al.}(2006)\citenamefont {Nadler}, \citenamefont {Lafon}, \citenamefont {Coifman},\ and\ \citenamefont {Kevrekidis}}]{nadlerDiffusionMapsSpectral2006}%
  \BibitemOpen
  \bibfield  {author} {\bibinfo {author} {\bibfnamefont {B.}~\bibnamefont {Nadler}}, \bibinfo {author} {\bibfnamefont {S.}~\bibnamefont {Lafon}}, \bibinfo {author} {\bibfnamefont {R.~R.}\ \bibnamefont {Coifman}},\ and\ \bibinfo {author} {\bibfnamefont {I.~G.}\ \bibnamefont {Kevrekidis}},\ }\bibfield  {title} {\bibinfo {title} {Diffusion maps, spectral clustering and reaction coordinates of dynamical systems},\ }\href {https://doi.org/10.1016/j.acha.2005.07.004} {\bibfield  {journal} {\bibinfo  {journal} {Applied and Computational Harmonic Analysis}\ }\bibinfo {series} {Special {{Issue}}: {{Diffusion Maps}} and {{Wavelets}}},\ \textbf {\bibinfo {volume} {21}},\ \bibinfo {pages} {113} (\bibinfo {year} {2006})}\BibitemShut {NoStop}%
\end{thebibliography}%
\end{document}